\documentclass[10pt,twocolumn,amsmath,amssymb,aps,pra,showpacs,longbibliography,superscriptaddress]{revtex4-1}
\usepackage[latin9]{inputenc}
\usepackage{amsmath}
\usepackage{amssymb}
\usepackage{graphicx}
\usepackage{esint}

\makeatletter

\newcommand*\LyXThinSpace{\,\hspace{0pt}}

\usepackage{amsfonts}
\usepackage{graphics}

\makeatother

\begin{document}
\title{Exact calculation of spectral properties of a particle interacting
with a one-dimensional Fermi gas in optical lattices}
\author{Xia-Ji Liu}
\email{Correspondence: xiajiliu@swin.edu.au}

\affiliation{Centre for Quantum Technology Theory, Swinburne University of Technology,
Melbourne 3122, Australia}
\author{Hui Hu}
\affiliation{Centre for Quantum Technology Theory, Swinburne University of Technology,
Melbourne 3122, Australia}
\date{\today}
\begin{abstract}
By using the exact Bethe wavefunctions of the one-dimensional Hubbard
model with $N$ spin-up fermions and one spin-down impurity, we derive
an analytic expression of the impurity form factor, in the form of
a determinant of a $(N+1)$ by $(N+1)$ matrix. This analytic expression
enables us to exactly calculate spectral properties of one-dimensional
Fermi polarons in lattices, when the masses of the impurity particle
and the Fermi bath are equal. We present the impurity spectral function
as functions of the on-site interaction strength and the filling factor
of the Fermi bath, and discuss the origin of Fermi singularities in
the spectral function at small momentum and the emergence of polaron
quasiparticles at large momentum near the boundary of Brillouin zone.
Our analytic expression of the impurity form factors pave the way
to exploring the intriguing dynamics of a particle interacting with
a Fermi bath. Our exact predictions on the impurity spectral function
could be directly examined in cold-atom laboratories by using the
radio-frequency spectroscopy and Ramsey spectroscopy.
\end{abstract}
\maketitle

\section{Introduction}

An impurity particle interacting with a Fermi bath of non-interacting
fermions is a traditional quantum many-body problem (namely, the Fermi
polaron problem) \citep{Alexandrov2010}, appearing in a diverse fields
of physics, including condensed matter physics, nuclear physics and
ultracold atomic physics. The research on Fermi polarons brings us
some fundamental concepts of quantum many-body physics, such as polaron
quasiparticles \citep{Landau1933}, Fermi-edge singularities \citep{Mahan1967,Nozieres1969}
and Anderson orthogonality catastrophe \citep{Anderson1967}. Over
the past two decades, the Fermi polaron problem receives a renewed
interest \citep{Massignan2014,Schmidt2018,Scazza2022,Wang2023AB,Tajima2024AB,Liu2024AB,Massignan2025},
due to the rapid experimental advances in ultracold atomic physics.
Due to the unprecedented controllability with ultracold atoms \citep{Bloch2008},
particularly in tuning the interparticle interaction strength \citep{Chin2010}
and in creating artificial optical lattices, various textbook models
of Fermi polarons can be experimentally realized \citep{Massignan2025}.
The spectral properties of Fermi polarons have now been quantitatively
measured, by using radio-frequency spectroscopy \citep{Schirotzek2009,Zhang2012,Yan2019},
Ramsey interferometry \citep{Cetina2016}, Rabi oscillation \citep{Scazza2017,Vivanco2024}
and Raman spectroscopy \citep{Ness2020}, and haven been theoretically
investigated, by using various many-body approaches, including variational
Chevy ansatz \citep{Chevy2006,Combescot2008,Parish2013,Liu2019,Hu2023AB,Hu2023ABpwave},
diagrammatic strong coupling theories \citep{Combescot2007,Hu2018,Tajima2019,Hu2022,Hu2024},
functional renormalization group \citep{Schmidt2011,vonMilczewski2024}
and quantum Monte Carlo simulations \citep{Prokofev2008,Goulko2016,Ramachandran2025}.

In spite of enormous efforts on studying the Fermi polaron, the quantitive
understanding of its spectral properties remains a grand challenge,
particularly at finite temperature \citep{Hu2024}. In this respect,
exact results of Fermi polarons in various limiting cases are highly
desirable. There are two well-known such cases. The first case is
the heavy polaron limit \citep{Nozieres1969,Schmidt2018,Knap2012,Wang2022PRL,Wang2022PRA},
where the impurity is infinitely heavy so it can be treated a scattering
potential. Another interesting non-trivial limit is the one dimension
case with equal mass for the impurity and the Fermi bath, where the
Fermi polaron problem can be exactly solved by using the celebrated
Bethe ansatz technique for the Gaudin-Yang model \citep{McGuire1966,Edwards1990,Castella1993,Guan2012}
and the one-dimensional (1D) Hubbard model \citep{Lieb1968,Deguchi2000,Essler2005}.
For a Fermi bath in free space as described by the Gaudin-Yang model,
the exact calculation of polaron spectral properties was presented
in the pioneering work by Castella and Zotos \citep{Castella1993}.
Most recently, the case of a 1D Fermi bath in optical lattices, as
described by the 1D Hubbard model, has also been discussed by the
present authors and collaborators, in the form of a short Letter \citep{Hu2025}.

In this work, we theoretically investigate the spectral properties
of Fermi polarons in 1D lattices, based on the exactly solvable 1D
Hubbard model \citep{Lieb1968,Deguchi2000,Essler2005}. The purpose
of our work is two-fold. On the one hand, we present the technical
details of the exact calculation reported earlier in the brief Letter
\citep{Hu2025}. In particular, we show in great detail how to derive
the impurity form factor, which is important for future studies on
the quantum dynamics of particles interacting with a Fermi bath \citep{Mathy2012,Knap2014,Gamayun2018,Dolgirev2021}.
On the other hand, we extend the theoretical calculations to the case
of attractive on-site interaction strength. We show that the two cases
of repulsion and attraction are related by a particle-hole transformation
for fermions in the Fermi bath. We also discuss the evolution of spectral
properties of Fermi polarons with increasing attractive interaction
and with varying the filling factor of the Fermi bath. These extended
results are useful for future experiments on Fermi polarons in 1D
optical lattices.

The rest of the paper is organized as follows. In the next section
(Sec. II), we briefly overview the Bethe ansatz solution of the 1D
Hubbard model with only one spin-down impurity. In Sec. III, we derive
the form factor of the regular Bethe wavefunctions. In Sec. IV, we
discuss the irregular quantum states that cannot be covered in the
Bethe ansatz solutions, such as the spin-flip states and the $\eta$-pairing
states \citep{Deguchi2000,Essler1992}. These irregular states are
necessary to include, as they make notable contributions to the impurity
spectral function at finite lattice sizes. In Sec. V, we present the
details of numerical calculations and emphasize how to select the
most important many-body states to exhaust the sum rule, as a useful
way to check the convergence of our numerical calculations. In Sec.
VI, we discuss the configuration of the many-body states that have
significant form factor or residue. We analyze in detail the spectral
properties of Fermi polarons at different filling factors and at attractive
on-site interaction strengths. Finally, Sec. VII is devoted to the
conclusions and outlooks.

\section{Bethe ansatz solution}

We start by considering a highly spin-population imbalanced spin-1/2
Fermi gas in 1D optical lattices with lattice size $L$, as described
by the standard 1D Hubbard model \citep{Lieb1968,Deguchi2000,Essler2005},
$\mathcal{H}=\mathcal{H}_{0}+\mathcal{H}_{U}$, where
\begin{eqnarray}
\mathcal{H}_{0} & = & -t\sum_{i=1}^{L}\sum_{\sigma=\uparrow,\downarrow}\left(\psi_{i\sigma}^{\dagger}\psi_{i+1\sigma}+h.c.\right),\\
\mathcal{H}_{U} & = & U\sum_{i=1}^{L}\psi_{i\uparrow}^{\dagger}\psi_{i\uparrow}\psi_{i\downarrow}^{\dagger}\psi_{i\downarrow},
\end{eqnarray}
are the non-interacting kinetic Hamiltonian and the interaction Hamiltonian
with strength $U$, respectively. We assume the periodic boundary
condition, so $\psi_{L+1\sigma}\equiv\psi_{1\sigma}$. This model
is exactly solvable by using the Bethe ansatz technique for any particle
numbers $N=\sum_{i}\psi_{i\uparrow}^{\dagger}\psi_{i\uparrow}$ and
$M=\sum_{i}\psi_{i\downarrow}^{\dagger}\psi_{i\downarrow}$. We focus
on the limit of a single spin-down fermion ($M=1$) and treat it as
an impurity. The filling factor of the lattice is then given by 
\begin{equation}
\nu\equiv\frac{N+M}{L}=\frac{N+1}{L}.
\end{equation}

\subsection{Bethe ansatz equations}

In this case, the many-body states $\left|\Psi_{N+1,Q}(\{k_{j,}\},\Lambda)\right\rangle $
of the system can be characterized by a set of the quasi-momenta $k_{j}$
($j=1,..,N+1$) and the quasi-momentum $\Lambda$, which satisfy $N+2$
coupled equations \citep{Lieb1968,Deguchi2000,Essler2005},
\begin{eqnarray}
\frac{\sin(k_{j})-\Lambda+iu}{\sin(k_{j})-\Lambda-iu} & = & e^{ik_{j}L},\\
\prod_{j=1}^{N+1}\frac{\sin(k_{j})-\Lambda+iu}{\sin(k_{j})-\Lambda-iu} & = & 1,
\end{eqnarray}
where $u=U/(4t)$ is the dimensionless interaction parameter. For
convenience, we assume that the lattice size $L$ is an even integer,
$N$ is an odd integer, and $L>N$. Once we solve $k_{j}$ for a particular
many-body state, the total momentum and energy of the state are obtained
by, 
\begin{equation}
Q=\sum_{j=1}^{N+1}k_{j}\;\textrm{mod}\;2\pi
\end{equation}
and 
\begin{equation}
E_{N+1}\left(\{k_{j}\},\Lambda\right)=-2t\sum_{j=1}^{N+1}\cos k_{j}
\end{equation}
respectively. We restrict the total momentum $Q$ to the first Brillouin
zone $(-\pi,+\pi]$ and always consider $Q\geq0$. We note that, for
a single impurity, the quasi-momentum $\Lambda$ is real. In contrast,
a complex-valued pair of the quasi-momenta (i.e., $k_{1}$ and $k_{2}=k_{1}^{*}$,
for concreteness) may arise for both attractive interaction ($u<0$)
and repulsive interaction ($u>0$), signifying a bound state. These
bound states are named as the $k-\Lambda$ string solutions. For convenience,
we refer to the other solutions with real quasi-momenta as the all
real-$k$ solutions.

\begin{figure}
\begin{centering}
\includegraphics[width=0.5\textwidth]{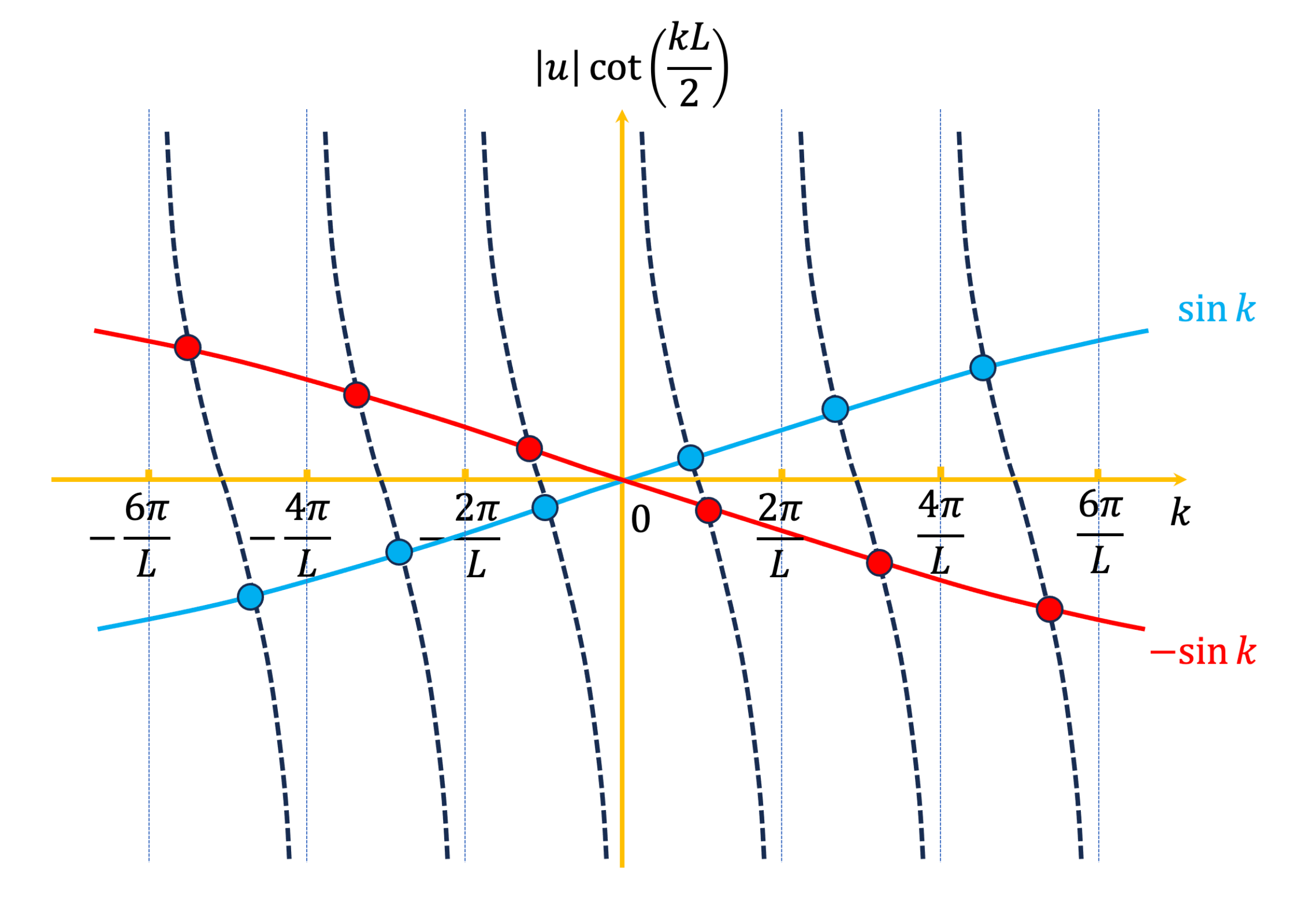}
\par\end{centering}
\caption{\label{fig1_BASolution} Graphical illustration of the solutions for
$\sin(k_{j})=u\cot(k_{j}L/2)$ in the ground state, where $\Lambda=0$.
The blue dots and red dots are the roots for the repulsive interaction
$u>0$ and the attractive interaction $u<0$, respectively.}
\end{figure}

It is useful to rewrite the coupled equations for $k_{j}$ and $\Lambda$
into an alternative form \citep{Deguchi2000},
\begin{eqnarray}
\sin\left(k_{j}\right)-\Lambda & = & u\cot\left(\frac{k_{j}L}{2}\right),\label{eq:BetheAnsatzRootEquation}\\
\sum_{j=1}^{N+1}k_{j} & = & n\frac{2\pi}{L},
\end{eqnarray}
where $-L/2<n\leq L/2$ is an integer. In the ground state, the total
momentum $Q=0$ and $n=0$, we must have $\Lambda=0$, as illustrated
in Fig. \ref{fig1_BASolution}. For a real quasi-momentum $k_{j}$,
for late convenience we introduce the variable \citep{Castella1993},
\begin{equation}
\delta_{j}=-\frac{\pi}{2}+\arctan\left[\frac{\sin\left(k_{j}\right)-\Lambda}{u}\right].
\end{equation}
According to the definition of $\delta_{j}$, it is easy to check
that
\begin{eqnarray}
e^{-2i\delta_{j}} & = & e^{ik_{j}L},\\
\frac{e^{i\delta_{j}}}{\sin\delta_{j}} & = & \frac{\sin\left(k_{j}\right)-\Lambda-iu}{-u}.
\end{eqnarray}
For the pair of the complex-valued quasi-momenta $k_{1}$ and $k_{2}$,
it is worth noting that we need to take care of the use of $\delta_{1}$
and $\delta_{2}$. In this case, actually it is more convenient to
directly use the complex-valued pair of $k_{1}$ and $k_{2}$.

\subsection{Impurity spectral function}

We are interested in calculating the impurity spectral function of
the single spin-down fermion. To this aim, we need to find out the
form factor of the many-body states $\left|\Psi_{N+1,Q}(\{k_{j}\},\Lambda)\right\rangle $,
defined by,
\begin{equation}
F_{N+1}\left(\{k_{j}\},\Lambda\right)=\left\langle \textrm{FS}_{N}\left|\psi_{Q\downarrow}\right|\Psi_{N+1,Q}\left(\{k_{j}\},\Lambda\right)\right\rangle ,
\end{equation}
which measures the overlap of the many-body state with the product
state $\psi_{Q\downarrow}^{\dagger}\left|\textrm{FS}_{N}\right\rangle $
(i.e., the product of a single-particle impurity state with momentum
$Q$ and a Fermi sea of $N$ non-interacting spin-up fermions). At
zero temperature, the impurity spectral function 
\begin{equation}
A(Q,\omega)=-\frac{1}{\pi}\textrm{Im}\mathcal{G}(Q,\omega)
\end{equation}
can be calculated from the Green function,
\begin{equation}
\mathcal{G}\left(Q,\omega\right)=\sum_{\{k_{j}\},\Lambda}\frac{Z_{N+1}\left(\{k_{j}\},\Lambda\right)}{\omega-\left[E_{N+1}\left(\{k_{j}\},\Lambda\right)-E_{\textrm{FS},N}\right]+i\delta},\label{eq:ApwN}
\end{equation}
where $E_{\textrm{FS},N}$ is the energy of the fully occupied Fermi
sea $\left|\textrm{FS}_{N}\right\rangle $ and 
\begin{equation}
Z_{N+1}(\{k_{j}\},\Lambda)\equiv\left|F_{N+1}(\{k_{j}\},\Lambda)\right|^{2}
\end{equation}
is the residue. For the finite lattice size $L$, we introduce a spectral
broadening factor $\delta\rightarrow0^{\dagger}$ to smooth the discrete
energy levels of many-body states. A typical choice is the average
single-particle spectral interval, $\delta=4t/L$, where $4t$ is
the band width of the 1D Hubbard model.

We note that, the summation in the Green function $\mathcal{G}(Q,\omega)$
is over all the many-body states of the system. These includes both
the regular Bethe wavefunctions $\left|\Psi_{N+1,Q}(\{k_{j}\},\Lambda)\right\rangle $
that can be constructed from the Bethe ansatz solutions with \emph{finite}
quasi-momenta $\{k_{j}\}$ and $\Lambda$ \citep{Deguchi2000}, and
some irregular states that we will discuss in detail in Sec. IV.

We note also that, the spectral function satisfies the sum-rule $\int_{-\infty}^{+\infty}d\omega A(Q,\omega)=1$,
as a result of the completeness of the many-body states, i.e., 
\begin{equation}
\sum_{\{k_{j}\},\Lambda}Z_{N+1}\left(\{k_{j}\},\Lambda\right)=1.\label{eq:sumrule}
\end{equation}
As we shall see, this sum-rule is very useful to check the convergence
of our numerical calculations.

\subsection{Particle-hole transformation}

The one-dimensional Hubbard model has an interesting particle-hole
transformation, which allows us to change the sign of the on-site
interaction strength \citep{Essler2005}. To see this, let us consider
the following transformation to the field operators $\psi_{i\uparrow}$
and $\psi_{i\uparrow}^{\dagger}$ for the Fermi bath of spin-up fermions,
$\psi_{i\uparrow}\rightarrow(-1)^{i}\psi_{i\uparrow}^{\dagger}$ and
$\psi_{i\uparrow}^{\dagger}\rightarrow(-1)^{i}\psi_{i\uparrow}$.
There is no change to the field operators for the spin-down impurity.
This is indeed a particle-hole transformation for the Fermi bath,
as the local density operator of spin-up fermions changes as,
\begin{equation}
\psi_{i\uparrow}^{\dagger}\psi_{i\uparrow}\rightarrow(-1)^{i}\psi_{i\uparrow}(-1)^{i}\psi_{i\uparrow}^{\dagger}=1-\psi_{i\uparrow}\psi_{i\uparrow}^{\dagger}.
\end{equation}
Therefore, in the thermodynamic limit ($L\rightarrow\infty$) the
filling factor changes from $\nu=\sum_{i}\psi_{i\uparrow}^{\dagger}\psi_{i\uparrow}/L$
to $\sum_{i}(1-\psi_{i\uparrow}^{\dagger}\psi_{i\uparrow})/L=1-\nu$. 

Under such a particle-hole transformation, the non-interacting Hamiltonian
is invariant, since the local hopping term $-t\psi_{i\uparrow}^{\dagger}\psi_{i+1\uparrow}+h.c.\rightarrow-t\left(-1\right)^{i}\psi_{i\uparrow}\left(-1\right)^{i+1}\psi_{i+1\uparrow}^{\dagger}+h.c.=-t\psi_{i+1\uparrow}^{\dagger}\psi_{i\uparrow}+h.c.$
does not change. Here, the minus sign due to the exchange of the fermionic
field operators is cancelled by the factor $(-1)^{i}(-1)^{i+1}=-1$.
On the other hand, we should have $\psi_{i\uparrow}^{\dagger}\psi_{i\uparrow}\psi_{i\downarrow}^{\dagger}\psi_{i\downarrow}\rightarrow(1-\psi_{i\uparrow}^{\dagger}\psi_{i\uparrow})\psi_{i\downarrow}^{\dagger}\psi_{i\downarrow}$
and hence,
\begin{equation}
\mathfrak{\mathcal{H}}_{U}\rightarrow U\sum_{i}(1-\psi_{i\uparrow}^{\dagger}\psi_{i\uparrow})\psi_{i\downarrow}^{\dagger}\psi_{i\downarrow}=\mathcal{H}_{-U}+U.
\end{equation}
It is clear that the on-site interaction Hamiltonian changes its sign
($U\rightarrow-U$). Additionally, we have an unimportant constant
energy shift $+U$. 

Therefore, with the particle-hole transformation for the spin-up Fermi
sea, the impurity problem with a positive on-site interaction $U>0$
at the filling factor $\nu$ is identical to the problem with an attractive
on-site interaction $-U<0$ at the filling factor $1-\nu$.

\section{The form factor}

We now derive the expression of the form factor $F_{N+1}\left(\{k_{j}\},\Lambda\right)$
for Bethe wavefunctions. In the first quantization, the real-space
wavefunction of the product state $\psi_{Q\downarrow}^{\dagger}\left|\textrm{FS}_{N}\right\rangle $
can be easily written down in terms of a $N\times N$ Slater determinant,
\begin{equation}
\psi_{Q\downarrow}^{\dagger}\left|\textrm{FS}_{N}\right\rangle =\frac{1}{\sqrt{N!}}\sqrt{\frac{1}{L}}e^{iQx_{\downarrow}}\det_{1\leq j,m\leq N}\left(\sqrt{\frac{1}{L}}e^{iq_{j}x_{m}}\right),
\end{equation}
where $x_{m}$ is the position of the $N$ spin-up fermions and 
\begin{equation}
q_{j}=\left[-\frac{N+1}{2}+j\right]\frac{2\pi}{L}
\end{equation}
is the associated momentum, which takes the values $[-(N-1)/2,\cdots,-1,0,1,\cdots,+(N-1)/2](2\pi/L)$
at zero temperature. In contrast, the expression of the Bethe ansatz
wavefunction $\left|\Psi_{N+1,Q}(\{k_{j}\},\Lambda)\right\rangle $
is much more involved. 

Let us start from the general form of the nested Bethe ansatz in the
region $R=\{x_{R0},\cdots x_{RN}\}$ with $x_{R0}<x_{R1}<\cdots<x_{R(N-1)}<x_{RN}$
\citep{Deguchi2000,Essler2005}:
\begin{equation}
\left|\Psi_{N+1,Q}(\{k_{j}\},\Lambda)\right\rangle \propto\sum_{P\in S_{N+1}}A\left(R;P\right)e{}^{i\sum_{j=0}^{N}k_{Pj}x_{Rj}}.\label{eq:BetheWF}
\end{equation}
Here, $R$ is a permutation of the $N+1$ coordinates $\{x_{\downarrow},x_{1},x_{2},\cdots,x_{N}\}$
and $P=\{P0,\cdots,PN\}$ is a permutation of the $N+1$ quasi-momenta
$\{k_{1},k_{2},\cdots,k_{N+1}\}$ in the permutation group $S_{N+1}$.
It is convenient to focus on the \emph{fundamental} region $R=\{x_{\downarrow}<x_{1}<x_{2}<\cdots<x_{N}\}$,
where the coefficients $A\left(R;P\right)$ takes the following form
(see the page 205 of Ref. \citep{Deguchi2000}),
\begin{align}
A\left(R;P\right) & =\epsilon\left(P\right)\prod_{j=1}^{N}\left[\frac{\sin\left(k_{Pj}\right)-\Lambda-iu}{-u}\right],\label{eq:AP}\\
 & =\epsilon\left(P\right)\prod_{j=1}^{N}\frac{e^{i\delta_{Pj}}}{\sin\delta_{Pj}},
\end{align}
up to an un-important normalization factor, which we shall determine
later. We have used $\epsilon(P)=\pm1$ to denote the parity of the
permutation $P$, so for an even permutation $\epsilon(P)=+1$ and
for an odd permutation $\epsilon(P)=-1$. 

\begin{widetext}The key idea of deriving the form factor is to rewrite
the Bethe wavefunction $\left|\Psi_{N+1,Q}(\{k_{j}\},\Lambda)\right\rangle $
into a $N\times N$ Slater determinant \citep{Edwards1990,Gamayun2015}.
Once we find it, the calculation of the form factor, i.e., the overlap
of two $N\times N$ Slater determinants, can be easily carried out
by using the celebrated identity \citep{Plasser2016},
\begin{equation}
\frac{1}{N!}\sum_{x_{1}=1}^{L}\cdots\sum_{x_{N}=1}^{L}\det_{1\leq j,m\leq N}\left[\varphi_{j}\left(x_{m}\right)\right]\det_{1\leq l,n\leq N}\left[\phi_{l}\left(x_{n}\right)\right]=\det_{1\leq j,l\leq N}\left[\sum_{x=1}^{L}\varphi_{j}\left(x\right)\phi_{l}\left(x\right)\right].\label{eq:FDA}
\end{equation}

\subsection{The Slater determinant representation of the Bethe wavefunction}

By substituting Eq. (\ref{eq:AP}) in the fundamental region into
the Bethe wavefunction Eq. (\ref{eq:BetheWF}), we obtain,
\begin{equation}
\left|\Psi_{N+1,Q}\{k_{j,}\Lambda\}\right\rangle \propto\sum_{P\in S_{N+1}}\epsilon\left(P\right)e^{ik_{P0}x_{\downarrow}}\prod_{j=1}^{N}\frac{\exp\left[ik_{Pj}x_{j}+i\delta_{Pj}\right]}{\sin\delta_{Pj}},
\end{equation}
where we explicitly single out the plane-wave function for the impurity
coordinate $x_{\downarrow}$. Let us now introduce the single-particle
wavefunction,
\begin{equation}
\chi_{j}\left(x\right)=\sqrt{\frac{1}{L}}\frac{\exp\left[ik_{j}x+i\delta_{j}\right]}{\sin\delta_{j}}=\left[\frac{\sin\left(k_{j}\right)-\Lambda-iu}{-u}\right]\sqrt{\frac{1}{L}}\exp\left[ik_{j}x\right].
\end{equation}
Therefore, the Bethe wavefunction $\left|\Psi_{N+1,Q}(\{k_{j}\},\Lambda)\right\rangle $
takes the form of a $(N+1)\times(N+1)$ determinant,
\begin{equation}
\left|\Psi_{N+1,Q}(\{k_{j}\},\Lambda)\right\rangle =\frac{C_{\Psi}}{\sqrt{N!L}}\sum_{P\in S_{N+1}}\epsilon\left(P\right)e^{ik_{P0}x_{\downarrow}}\prod_{j=1}^{N}\chi_{Pj}\left(x_{j}\right)=\frac{C_{\Psi}}{\sqrt{N!L}}\left|\begin{array}{cccc}
e^{ik_{1}x_{\downarrow}} & \chi_{1}\left(x_{1}\right) & \cdots & \chi_{1}\left(x_{N}\right)\\
\vdots & \ddots & \cdots & \vdots\\
\vdots & \vdots & \ddots & \vdots\\
e^{ik_{N+1}x_{\downarrow}} & \chi_{N+1}\left(x_{1}\right) & \cdots & \chi_{N+1}\left(x_{N}\right)
\end{array}\right|,
\end{equation}
where $C_{\Psi}$ is the normalization factor to be determined soon.
As we work in the fundamental region $R=\{x_{\downarrow}<x_{1}<x_{2}<\cdots<x_{N}\}$,
it is useful to introduce new coordinates $y_{j}=x_{j}-x_{\downarrow}$
relative to the impurity position and rewrite the wavefunction into
the form,
\begin{align}
\left|\Psi_{N+1,Q}(\{k_{j}\},\Lambda)\right\rangle  & =\frac{C_{\Psi}}{\sqrt{N!}}\sqrt{\frac{1}{L}}e^{iQx_{\downarrow}}\left|\begin{array}{cccc}
1 & \chi_{1}\left(y_{1}\right) & \cdots & \chi_{1}\left(y_{N}\right)\\
\vdots & \ddots & \cdots & \vdots\\
\vdots & \vdots & \ddots & \vdots\\
1 & \chi_{N+1}\left(y_{1}\right) & \cdots & \chi_{N+1}\left(y_{N}\right)
\end{array}\right|,\\
 & =\frac{C_{\Psi}}{\sqrt{N!}}\sqrt{\frac{1}{L}}e^{iQx_{\downarrow}}\left|\begin{array}{ccc}
\chi_{1}\left(y_{1}\right)-\chi_{N+1}\left(y_{1}\right) & \cdots & \chi_{1}\left(y_{N}\right)-\chi_{N+1}\left(y_{N}\right)\\
\ddots & \cdots & \vdots\\
\chi_{N}\left(y_{1}\right)-\chi_{N+1}\left(y_{1}\right) & \ddots & \chi_{N}\left(y_{N}\right)-\chi_{N+1}\left(y_{N}\right)
\end{array}\right|,
\end{align}
where in the second expression, we have absorbed a trivial factor
of $(-1)^{N}=-1$ into the normalization factor $C_{\Psi}$. To further
simplify the notation, let us introduce another single-particle wavefunction,
\begin{equation}
\phi_{j}\left(y\right)=\chi_{j}\left(y\right)-\chi_{N+1}\left(y\right)=\sqrt{\frac{1}{L}}\frac{\exp\left[ik_{j}y+i\delta_{j}\right]}{\sin\delta_{j}}-\sqrt{\frac{1}{L}}\frac{\exp\left[ik_{N+1}y+i\delta_{N+1}\right]}{\sin\delta_{N+1}}.\label{eq:faij}
\end{equation}
We finally cast the Bethe wavefunction into the desired $N\times N$
Slater determinant,
\begin{equation}
\left|\Psi_{N+1,Q}\{k_{j,}\Lambda\}\right\rangle =\frac{C_{\Psi}}{\sqrt{N!}}\sqrt{\frac{1}{L}}e^{iQx_{\downarrow}}\det_{1\leq j,m\leq N}\left[\phi_{j}\left(y_{m}\right)\right].
\end{equation}

\subsection{The normalization factor of the Bethe wavefunction}

The normalization factor $C_{\Psi}$ can be easily determined by using
the identity Eq. (\ref{eq:FDA}). By applying the normalization $\left\langle \Psi_{N+1,Q}(\{k_{j}\},\Lambda)\mid\Psi_{N+1,Q}(\{k_{j}\},\Lambda)\right\rangle =1$,
after completing the trivial summation on $x_{\downarrow}$, it is
straightforward to obtain
\begin{equation}
\left|C_{\Psi}\right|^{-2}=\frac{1}{N!}\sum_{y_{1}=1}^{L}\cdots\sum_{y_{N}=1}^{L}\det_{1\leq j,m\leq N}\left[\phi_{j}^{*}\left(y_{m}\right)\right]\det_{1\leq l,n\leq N}\left[\phi_{l}\left(y_{n}\right)\right]=\det_{1\leq j,l\leq N}\left[\sum_{y=1}^{L}\phi_{j}^{*}\left(y\right)\phi_{l}\left(y\right)\right].
\end{equation}
To calculate the summation $\sum_{y=1}^{L}\phi_{j}^{*}(y)\phi_{l}(y)$,
let us first check 
\begin{equation}
A_{jl}=\sum_{y=1}^{L}\bar{\chi}_{j}\left(y\right)\chi_{l}\left(y\right),
\end{equation}
where $j,l=1,\cdots,N+1$ and 
\begin{equation}
\bar{\chi}_{j}\left(y\right)\equiv\left[\frac{\sin\left(k_{j}\right)-\Lambda+iu}{-u}\right]\sqrt{\frac{1}{L}}\exp\left[-ik_{j}y\right].
\end{equation}
Note that, for a real quasi-momentum $k_{j}$, we trivially have $\chi_{j}^{*}(y)=\bar{\chi}_{j}(y)$.
We find that,
\begin{equation}
A_{jl}=\left[\frac{\sin\left(k_{j}\right)-\Lambda+iu}{-u}\right]\left[\frac{\sin\left(k_{l}\right)-\Lambda-iu}{-u}\right]\frac{1}{L}\sum_{y=1}^{L}\exp\left[-i\left(k_{j}-k_{l}\right)y\right].
\end{equation}
Let us consider the case with $j\neq l$:
\begin{eqnarray}
A_{jl} & = & \left[\frac{\sin\left(k_{j}\right)-\Lambda+iu}{-u}\right]\left[\frac{\sin\left(k_{l}\right)-\Lambda-iu}{-u}\right]\frac{1}{L}\frac{e^{-i\left(k_{j}-k_{l}\right)}}{1-e^{-i\left(k_{j}-k_{l}\right)}}\left[1-e^{-i\left(k_{j}-k_{l}\right)L}\right],\\
 & = & \frac{2i\left[\sin\left(k_{l}\right)-\sin\left(k_{j}\right)\right]}{u}\frac{1}{L}\frac{1}{e^{i\left(k_{j}-k_{l}\right)}-1},\\
 & = & -\frac{e^{-ik_{j}}+e^{ik_{l}}}{Lu}.
\end{eqnarray}
Here, in the first step we have used the Bethe ansatz equation such
as $e^{ik_{j}L}=[\sin(k_{j})-\Lambda+iu]/[\sin(k_{j})-\Lambda-iu]$
and in the second step we have used the identity $\left[e^{i(k_{j}-k_{l})}-1\right](e^{-ik_{j}}+e^{ik_{l}})=-2i[\sin(k_{l})-\sin(k_{j})]$.
For the case with $j=l$, we obtain
\begin{equation}
A_{jj}=\left[\frac{\sin\left(k_{j}\right)-\Lambda+iu}{-u}\right]\left[\frac{\sin\left(k_{j}\right)-\Lambda-iu}{-u}\right]=\frac{1}{\sin^{2}\delta_{j}},
\end{equation}
where the second expression in terms of $\sin\delta_{j}$ is convenient
for a \emph{real} quasi-momentum $k_{j}$. 

For simplicity, let us now assume that all the quasi-momenta are real.
The complex-valued pair of quasi-momenta can be easily treated in
a moment. By using $A_{jl}$, it is easy to see that ($j,l=1,\cdots,N$),
\begin{eqnarray}
\sum_{y=1}^{L}\phi_{j}^{*}\left(y\right)\phi_{l}\left(y\right) & = & \sum_{y=1}^{L}\left[\bar{\chi}_{j}\left(y\right)-\bar{\chi}_{N+1}\left(y\right)\right]\left[\chi_{l}\left(y\right)-\chi_{N+1}\left(y\right)\right],\\
 & = & A_{jl}+\frac{e^{-ik_{j}}+e^{ik_{N+1}}}{Lu}+\frac{e^{-ik_{N+1}}+e^{ik_{l}}}{Lu}+\frac{1}{\sin^{2}\delta_{N+1}}.
\end{eqnarray}
At this point, it is useful to define the variable 
\begin{equation}
u_{j}=\frac{1}{\sin^{2}\delta_{j}}+\frac{2\cos\left(k_{j}\right)}{Lu}=1+\frac{\left[\sin\left(k_{j}\right)-\Lambda\right]^{2}}{u^{2}}+\frac{2\cos\left(k_{j}\right)}{Lu},\label{eq:uj}
\end{equation}
where the second expression is more general and holds for a complex-valued
quasi-momentum $k_{j}$. Then, it is straightforward to obtain,
\begin{equation}
\sum_{y=1}^{L}\phi_{j}^{*}\left(y\right)\phi_{l}\left(y\right)=\left\{ \begin{array}{cc}
u_{j}+u_{N+1} & \textrm{if }j=l\\
u_{N+1} & \textrm{if }j\neq l
\end{array}\right..
\end{equation}
The expression of the normalization factor $C_{\Psi}$ becomes,
\begin{equation}
\left|C_{\Psi}\right|^{-2}=\left|\begin{array}{cccc}
u_{1}+u_{N+1} & u_{N+1} & \cdots & u_{N+1}\\
u_{N+1} & u_{2}+u_{N+1} & \cdots & u_{N+1}\\
\vdots & \vdots & \ddots & \vdots\\
u_{N+1} & u_{N+1} & \cdots & u_{N}+u_{N+1}
\end{array}\right|=\left|\begin{array}{ccccc}
u_{1} & -u_{2} & 0 & \cdots & 0\\
0 & u_{2} & -u_{3} & \cdots & 0\\
\vdots & \ddots & \ddots & \cdots & \vdots\\
0 & 0 & \cdots & u_{N-1} & -u_{N}\\
u_{N+1} & u_{N+1} & \cdots & u_{N+1} & u_{N}+u_{N+1}
\end{array}\right|.
\end{equation}
This $N\times N$ determinant is easy to calculate in a recursive
way and we find that,
\begin{equation}
\left|C_{\Psi}\right|^{-2}=u_{1}u_{2}\cdots u_{N+1}\left(\frac{1}{u_{1}}+\frac{1}{u_{2}}+\cdots+\frac{1}{u_{N+1}}\right),
\end{equation}
which is a symmetric function of $u_{1},u_{2},\cdots,u_{N+1}$, as
expected. 

In the case of a (single) complex-valued pair of quasi-momenta, let
us assume for concreteness that they are given by $k_{1}=k_{R}+ik_{I}$
and $k_{2}=k_{R}-ik_{I}$. We only need to take care of $\chi_{1}^{*}(y)=\bar{\chi}_{2}(y)$
and $\chi_{2}^{*}(y)=\bar{\chi}_{1}(y)$, i.e., the row index $1$
and $2$ are interchanged. As a result, we need to exchange the first
row and second row of the previous $N\times N$ determinant for $\left|C_{\Psi}\right|^{-2}$.
This brings a \emph{minus} sign for $\left|C_{\Psi}\right|^{-2}$.
Let us denote the number of pairs of complex-valued quasi-momenta
in the Bethe ansatz solution by an integer variable $I_{B}$, although
we only have one pair at most in our Fermi polaron problem. Then,
we may write the general expression,
\begin{equation}
\left|C_{\Psi}\right|^{-2}=\left(-1\right)^{I_{B}}u_{1}u_{2}\cdots u_{N+1}\left(\frac{1}{u_{1}}+\frac{1}{u_{2}}+\cdots+\frac{1}{u_{N+1}}\right),\label{eq:CFAI}
\end{equation}
where $I_{B}=0$ and $I_{B}=1$ for the all real-$k$ states and the
$k-\Lambda$ string states, respectively. We note that, in general,
$u_{j}$ could be complex-valued due to the bound-state and would
be better calculated in terms of $k_{j}$ in Eq. (\ref{eq:uj}). We
note also that, Eq. (\ref{eq:CFAI}) is valid for $N=1$. In this
case, we have only one single-particle state $\phi_{1}(y)=\chi_{1}(y)-\chi_{2}(y)$.
For the complex-value pair, we then have $\phi_{1}^{*}(y)=\chi_{1}^{*}(y)-\chi_{2}^{*}(y)=-[\bar{\chi}_{1}(y)-\bar{\chi}_{2}(y)]$.
A minus sign automatically appears.

\subsection{The form factor}

We are now ready to derive the expression of the form factor. Let
us slightly rewrite the product wavefunction in terms of the relative
coordinates $y_{m}$ ($m=1,\cdots,N$),
\begin{equation}
\psi_{Q\downarrow}^{\dagger}\left|\textrm{FS}_{N}\right\rangle =\frac{1}{\sqrt{N!}}\sqrt{\frac{1}{L}}e^{iQx_{\downarrow}}\det_{1\leq j,m\leq N}\left(\sqrt{\frac{1}{L}}e^{iq_{j}y_{m}}\right).
\end{equation}
The change of $x_{m}$ to $y_{m}$ is allowed, since $\sum_{j}q_{j}x_{\downarrow}=0$.
Therefore, by using the identity Eq. (\ref{eq:FDA}), the form factor
$F_{N+1}(\{k_{j}\},\Lambda)$ is given by
\begin{align}
F_{N+1}\left(\{k_{j}\},\Lambda\right) & =\frac{C_{\Psi}}{N!}\sum_{y_{1}=1}^{L}\cdots\sum_{y_{N}=1}^{L}\det_{1\leq j,m\leq N}\left(\sqrt{\frac{1}{L}}e^{-iq_{j}y_{m}}\right)\det_{1\leq l,n\leq N}\left[\phi_{l}\left(y_{n}\right)\right],\\
 & =C_{\Psi}\det_{1\leq j,l\leq N}\left[\sqrt{\frac{1}{L}}\sum_{y=1}^{L}e^{-iq_{j}y}\phi_{l}\left(y\right)\right].
\end{align}
To complete the summation in the determinant, let us first check
\begin{eqnarray}
B_{jl} & = & \sqrt{\frac{1}{L}}\sum_{y=1}^{L}e^{-iq_{j}y}\chi_{l}\left(y\right)=\left[\frac{\sin\left(k_{l}\right)-\Lambda-iu}{-u}\right]\frac{1}{L}\sum_{y=1}^{L}\exp\left[-i\left(q_{j}-k_{l}\right)y\right],\\
 & = & \left[\frac{\sin\left(k_{l}\right)-\Lambda-iu}{-u}\right]\frac{1}{L}\frac{e^{-i\left(q_{j}-k_{l}\right)}}{1-e^{-i\left(q_{j}-k_{l}\right)}}\left[1-e^{-i\left(q_{j}-k_{l}\right)L}\right].
\end{eqnarray}
By using $e^{-iq_{j}L}=1$ and $e^{ik_{l}L}=[\sin(k_{l})-\Lambda+iu]/[\sin(k_{l})-\Lambda-iu]$,
we find that, 
\begin{equation}
B_{jl}=\frac{2i}{L}\frac{1}{e^{i\left(q_{j}-k_{l}\right)}-1}=\frac{1}{L}\frac{e^{-iq_{j}}+e^{ik_{l}}}{\sin\left(q_{j}\right)-\sin\left(k_{l}\right)}.
\end{equation}
Therefore, we obtain,
\begin{equation}
\sqrt{\frac{1}{L}}\sum_{y=1}^{L}e^{-iq_{j}y}\phi_{l}\left(y\right)=B_{jl}-B_{j,N+1}=\frac{1}{L}\frac{e^{-iq_{j}}+e^{ik_{l}}}{\sin\left(q_{j}\right)-\sin\left(k_{l}\right)}-\frac{1}{L}\frac{e^{-iq_{j}}+e^{ik_{N+1}}}{\sin\left(q_{j}\right)-\sin\left(k_{N+1}\right)}.
\end{equation}
Finally, we may write the form factor in terms of a $(N+1)\times(N+1)$
determinant,
\begin{equation}
F_{N+1}\left(\{k_{j}\},\Lambda\right)=C_{\Psi}\left|\begin{array}{cccc}
B_{11} & \cdots & B_{1N} & B_{1,N+1}\\
\vdots & \vdots & \vdots & \vdots\\
B_{N1} & \cdots & B_{NN} & B_{N,N+1}\\
1 & \cdots & 1 & 1
\end{array}\right|.\label{eq:FormFactor}
\end{equation}
\end{widetext}

\section{the irregular many-body states}

As we mentioned earlier, the regular Bethe wavefunctions with finite
quasi-momenta $\{k_{j}\}$ and $\Lambda$ do not cover all the many-body
states of the one-dimensional Hubbard model \citep{Deguchi2000,Essler1992}.
We have other irregular quantum states, which contribute to the important
sum rule Eq. (\ref{eq:sumrule}). This is clearly seen in numerical
calculations. While at zero or small total momentum, the sum of the
residues of all the regular Bethe wavefunctions is exactly unity,
at large total momentum (i.e., $Q=\pi$) the sum is less than 1, due
to the exclusion of irregular quantum states that have significant
residue.

\subsection{A simple example with two-site Hubbard model}

A simple example of irregular quantum states can be obtained by considering
a two-site Hubbard model ($L=2$). As discussed in Ref. \citep{Essler1992},
in the $(1+1)$-sector of one spin-up fermion and one spin-down fermion,
we have the following two regular Bethe states with a total momentum
$Q=0$ and the quasi-momentum $\Lambda=0$, and with the energies
$D_{\mp}=[U/2\mp\sqrt{U^{2}/4+16t^{2}}]/t$,
\begin{eqnarray}
\psi_{\downarrow\uparrow}\left(k_{1}^{-},k_{2}^{-}\right) & = & \frac{1}{\sqrt{D_{-}^{2}/8+2}}\left[-\frac{D_{-}}{4}\left|\psi_{1}\right\rangle +\left|\psi_{2}\right\rangle \right],\\
\psi_{\downarrow\uparrow}\left(k_{1}^{+},k_{2}^{+}\right) & = & \frac{1}{\sqrt{D_{+}^{2}/8+2}}\left[-\frac{D_{+}}{4}\left|\psi_{1}\right\rangle +\left|\psi_{2}\right\rangle \right],
\end{eqnarray}
where 
\begin{eqnarray}
\left|\psi_{1}\right\rangle  & = & \left(\psi_{1\uparrow}^{\dagger}\psi_{1\downarrow}^{\dagger}+\psi_{2\uparrow}^{\dagger}\psi_{2\downarrow}^{\dagger}\right)\left|0\right\rangle ,\\
\left|\psi_{2}\right\rangle  & = & \left(\psi_{1\uparrow}^{\dagger}\psi_{2\downarrow}^{\dagger}-\psi_{1\downarrow}^{\dagger}\psi_{2\uparrow}^{\dagger}\right)\left|0\right\rangle ,\\
e^{ik_{1}^{\pm}} & = & -\frac{D_{\pm}}{4}+\sqrt{\left(\frac{D_{\pm}}{4}\right)^{2}-1}.
\end{eqnarray}
The total residues of these two regular Bethe states are precisely
unity at the total momentum $Q=0$, as numerically confirmed. 

As the dimension of the total Hilbert space in the $(1+1)$-sector
is $4$ (i.e., the spin-up fermion and the spin-down fermion each
can occupy two sites), we know that there are two other quantum states.
It is readily seen that the two states are given by, 
\begin{eqnarray}
\left|\eta\right\rangle  & = & \frac{1}{\sqrt{2}}\left(\psi_{1\uparrow}^{\dagger}\psi_{1\downarrow}^{\dagger}-\psi_{2\uparrow}^{\dagger}\psi_{2\downarrow}^{\dagger}\right)\left|0\right\rangle ,\\
\left|\zeta\right\rangle  & = & \frac{1}{\sqrt{2}}\left(\psi_{1\uparrow}^{\dagger}\psi_{2\downarrow}^{\dagger}+\psi_{1\downarrow}^{\dagger}\psi_{2\uparrow}^{\dagger}\right)\left|0\right\rangle ,
\end{eqnarray}
and have a total momentum $Q=\pi=2\pi/L$. Both $\left|\eta\right\rangle $
and $\left|\zeta\right\rangle $ states are \emph{not} described by
the regular Bethe wavefunctions, i.e., in Eq. (\ref{eq:BetheAnsatzRootEquation})
we can not find finite quasi-momenta $k_{1}$, $k_{2}$ and $\Lambda$,
that can reproduce these two states. At the total momentum $Q=\pi$,
we find that the states $\left|\eta\right\rangle $ and $\left|\zeta\right\rangle $
have the form factor $1/\sqrt{2}$ and $-1/\sqrt{2}$, respectively,
and they exhaust the sum rule Eq. (\ref{eq:sumrule}).

\subsection{The construction of irregular quantum states}

Quite generally, it turns out that the irregular quantum states could
be generated by acting the $\zeta^{\dagger}$ spin-flip operator and
$\eta^{\dagger}$ pairing operator to some quantum states $\left|\varphi\right\rangle $.
These two operators are defined as follows,
\begin{eqnarray}
\zeta & = & \sum_{j=1}^{L}\psi_{j\uparrow}^{\dagger}\psi_{j\downarrow}=\sum_{k}\psi_{k\uparrow}^{\dagger}\psi_{k\downarrow},\\
\eta & = & \sum_{j=1}^{L}\left(-1\right)^{j}\psi_{j\uparrow}\psi_{j\downarrow}=\sum_{k}\psi_{\pi-k\uparrow}\psi_{k\downarrow}.
\end{eqnarray}
As both $\zeta^{\dagger}$ and $\eta^{\dagger}$ contain a creation
operator for the impurity particle, it is easy to see that the quantum
states $\left|\varphi\right\rangle $ should be the non-interacting
states consisting of either $N+1$ or $N-1$ spin-up fermions, which
are very easy to construct (see below for the construction). Moreover,
the states $\zeta^{\dagger}\left|\varphi\right\rangle $ and $\left|\varphi\right\rangle $
have the same energy (i.e., $E_{\varphi}$), so the action of the
spin-flip operator $\zeta^{\dagger}$ does not change the energy of
the state. In contrast, the action of the $\eta$-pairing operator
$\eta^{\dagger}$ will increase the energy by an amount $U$. To see
this, let us recall the commutation relation \citep{Yang1989},
\begin{equation}
\left[\eta^{\dagger},\mathcal{H}\right]=\eta^{\dagger}\mathcal{H-H\eta^{\dagger}=}-U\eta^{\dagger}.
\end{equation}
It is readily seen that, 
\begin{equation}
\mathcal{H}\eta^{\dagger}\left|\varphi\right\rangle =\left(E_{\varphi}+U\right)\eta^{\dagger}\left|\varphi\right\rangle ,
\end{equation}
showing that the state $\eta^{\dagger}\left|\varphi\right\rangle $
is an exact eigenstate of the Hamiltonian that has an energy $U$
more than the non-interacting state $\left|\varphi\right\rangle $
\citep{Yang1989}.

We now wish to construct the irregular quantum states $\zeta^{\dagger}\left|\varphi\right\rangle $
and $\eta^{\dagger}\left|\varphi\right\rangle $ that have non-zero
overlap with the initial product state, 
\begin{equation}
\left|\Psi_{0}\right\rangle =\psi_{Q\downarrow}^{\dagger}\left|\textrm{FS}_{N}\right\rangle =\psi_{Q\downarrow}^{\dagger}\prod_{j=1}^{N}\psi_{q_{j}\uparrow}^{\dagger}\left|0\right\rangle ,
\end{equation}
where $q_{j}\in[-(N-1)/2,\cdots,-1,0,1,\cdots,+(N-1)/2](2\pi/L)$.
For this purpose, let us consider the states, $\zeta\left|\Psi_{0}\right\rangle $
and $\eta\left|\Psi_{0}\right\rangle $,\begin{widetext}
\begin{eqnarray}
\zeta\left|\Psi_{0}\right\rangle  & = & \left(\sum_{k}\psi_{k\uparrow}^{\dagger}\psi_{k\downarrow}\right)\psi_{Q\downarrow}^{\dagger}\prod_{j=1}^{N}\psi_{q_{j}\uparrow}^{\dagger}\left|0\right\rangle =\psi_{Q\uparrow}^{\dagger}\prod_{j=1}^{N}\psi_{q_{j}\uparrow}^{\dagger}\left|0\right\rangle ,\\
\eta\left|\Psi_{0}\right\rangle  & = & \left(\sum_{k}\psi_{\pi-k\uparrow}\psi_{k\downarrow}\right)\psi_{Q\downarrow}^{\dagger}\prod_{j=1}^{N}\psi_{q_{j}\uparrow}^{\dagger}\left|0\right\rangle =\psi_{\pi-Q\uparrow}\prod_{j=1}^{N}\psi_{q_{j}\uparrow}^{\dagger}\left|0\right\rangle .
\end{eqnarray}
\end{widetext}It is easy to see that the state $\zeta\left|\Psi_{0}\right\rangle $
exists if $Q\notin\{q_{j}\}$, and the state $\eta\left|\Psi_{0}\right\rangle $
exists if $\pi-Q\in\{q_{j}\}$. Let us discuss the two cases separately.

\subsubsection{The $\zeta$ spin-flip state}

In the case that $Q\notin\{q_{j}\}$, we may construct a $\zeta$
spin-flip state, 
\begin{align}
\left|\Psi_{\zeta}\right\rangle  & =C_{\zeta}\zeta^{\dagger}\left[\psi_{Q\uparrow}^{\dagger}\prod_{j=1}^{N}\psi_{q_{j}\uparrow}^{\dagger}\left|0\right\rangle \right],\\
 & =\frac{\zeta^{\dagger}}{\sqrt{N+1}}\left[\psi_{Q\uparrow}^{\dagger}\prod_{j=1}^{N}\psi_{q_{j}\uparrow}^{\dagger}\left|0\right\rangle \right].
\end{align}
Here, the normalization coefficient $C_{\zeta}=1/\sqrt{N+1}$, since
$N+1$ orthogonal terms are created, when we act the spin-flip operator
$\zeta^{\dagger}$ on $\left|\varphi\right\rangle =\zeta\left|\Psi_{0}\right\rangle $.
The energy of this $\zeta$ spin-flip state is given by $E_{0}=-2t\cos Q+E_{\textrm{FS},N}$,
where $E_{\textrm{FS},N}$ is the energy of the Fermi sea, as we defined
earlier. Its overlap or the form factor with the initial state $\left|\Psi_{0}\right\rangle $
is simply $F_{\zeta}=C_{\zeta}=1/\sqrt{N+1}$.

\subsubsection{The $\eta$-pairing state}

In the case that $\pi-Q\in\{q_{j}\}$, we may similarly construct
an $\eta$-pairing state, 
\begin{align}
\left|\Psi_{\eta}\right\rangle  & =C_{\eta}\eta^{\dagger}\left[\psi_{\pi-Q\uparrow}\prod_{j=1}^{N}\psi_{q_{j}\uparrow}^{\dagger}\left|0\right\rangle \right],\\
 & =\frac{\eta^{\dagger}}{\sqrt{L-N+1}}\left[\psi_{\pi-Q\uparrow}\prod_{j=1}^{N}\psi_{q_{j}\uparrow}^{\dagger}\left|0\right\rangle \right].
\end{align}
The normalization coefficient $C_{\eta}=1/\sqrt{L-N+1}$, because
$L-N+1$ orthogonal terms are generated, when we act the $\eta$-pairing
operator $\eta^{\dagger}$ on $\left|\varphi\right\rangle =\eta\left|\Psi_{0}\right\rangle $.
It is easy to see that the energy of the state $\eta\left|\Psi_{0}\right\rangle $
is also $E_{0}=-2t\cos Q+E_{\textrm{FS},N}$, since in this state
we \emph{remove} from the Fermi sea a single-particle state with momentum
$\pi-Q$ with energy $-2t\cos(\pi-Q)=2t\cos Q$. Therefore, the energy
of the $\eta$-paring state $\left|\Psi_{\eta}\right\rangle $ is
$E_{0}+U$. Its overlap or the form factor with the initial state
$\left|\Psi_{0}\right\rangle $ is given by $F_{\eta}=C_{\eta}=1/\sqrt{L-N+1}$.

\subsection{The sum-rule}

We can now work out the sum-rule for the sum $\varrho_{\textrm{BA}}$
of the residues of all the regular Bethe wavefunctions and compare
$\varrho_{\textrm{BA}}$ with the sum of all the residues (i.e., $\varrho_{s}=1$).
Let us define the Fermi wavevector $k_{F}=N(2\pi/L)$, $k_{c1}=\min\{k_{F},\pi-k_{F}\}$
and $k_{c2}=\max\{k_{F},\pi-k_{F}\}$. For the momentum $Q\in[0,\pi]$,
we find that,
\begin{equation}
\varrho_{\textrm{BA}}=\begin{cases}
\begin{array}{c}
1,\\
1-F_{\zeta}^{2},\\
1-F_{\zeta}^{2}-F_{\eta}^{2},
\end{array} & \begin{array}{c}
Q<k_{c1}\\
k_{c1}<Q<k_{c2}\\
k_{c2}<Q
\end{array}\end{cases}.
\end{equation}
As the special cases, $\varrho_{\textrm{BA}}=1$ at $Q=0$ and $\varrho_{\textrm{BA}}=1-(L+2)/[(N+1)(L-N+1)]$
at $Q=\pi$. In the thermodynamic limit ($N\rightarrow\infty$ and
$L\rightarrow\infty$), we always have $\varrho_{\textrm{BA}}\rightarrow\varrho_{s}=1$,
as expected.

For later convenience, we define the sum of the residues from the
irregular quantum states as $\varrho_{\textrm{irr}}$, which is $\varrho_{\textrm{irr}}=F_{\zeta}^{2}=1/(N+1)$
when $k_{c1}<Q<k_{c2}$ and is $\varrho_{\textrm{irr}}=F_{\zeta}^{2}+F_{\eta}^{2}=1/(N+1)+1/(L-N+1)$
when $k_{c2}<Q$. We also denote the sums of the residues from the
all real-$k$ solutions (including the spin-flip state) and from the
$k-\Lambda$ string solutions (including the $\eta$-pairing state)
as $\varrho_{k}$ and $\varrho_{\Lambda}$, respectively. Therefore,
$\varrho_{k}+\varrho_{\Lambda}=1$. 

\section{Numerical calculations}

For a given many-body state with a set of quasi-momenta $\{k_{j}\}$
and $\Lambda$, the calculation of its form factor in Eq. (\ref{eq:FormFactor})
is straightforward. We first calculate $u_{j}$ in Eq. (\ref{eq:uj})
by using $k_{j}$ and determine the normalization factor of $C_{\Psi}$
in Eq. (\ref{eq:CFAI}). We then calculate the $(N+1)\times(N+1)$
determinant in Eq. (\ref{eq:FormFactor}) to obtain the form factor.
We need to pay special attention for the $k-\Lambda$ string solutions
with a pair of complex-valued quasi-momenta, particularly for the
normalization factor $C_{\Psi}$. For convenience, we always consider
the case that the lattice size $L$ is an even integer.

\subsection{The numerical solution of the Bethe ansatz equations}

From Fig. \ref{fig1_BASolution}, we find that each quasi-momentum
$k_{j}$ is uniquely characterized by an integer $s_{j}$ in the region
$(s_{j}-1)2\pi/L\leq k_{j}<s_{j}2\pi/L$. As we restrict the quasi-momentum
in the first Brillouin zone $k\in(-\pi,+\pi]$, the allowed values
for $s_{j}$ range from $-L/2+1$ to $L/2$. For the solution with
all real momenta $\{k_{j}\}$, we simply assign the different integers
$\{s_{j}\}$ ($j=1,\cdots,N+1$) and the integer $n=Q/(2\pi/L)$,
as a set of good quantum numbers to characterize the regular Bethe
ansatz solution. We note that, in the standard textbook the quasi-momentum
$k_{j}$ is often specified by a half integer $I_{j}$ for an odd
number of particles $N$ \citep{Deguchi2000,Essler2005,Kohno2010}.
Our integer quantum number $s_{j}$ is related to $I_{j}$ by $s_{j}=I_{j}+1/2$.

Numerically, for a selected quasi-momentum $\Lambda\in(-\infty,+\infty)$,
we can solve Eq. (\ref{eq:BetheAnsatzRootEquation}) for $\{k_{j}\}$,
once a set of integers $\{s_{j}\}$ is specified. We then obtain the
total momentum $Q=\sum_{j=1}^{N+1}k_{j}=n(2\pi/L)$. As we shall see
below for the cases of $U<0$ and $U>0$ separately, there is one-to-one
correspondence between the quasi-momentum $\Lambda$ and the total
momentum $Q$. Therefore, we can adjust the value of the quasi-momentum
$\Lambda$ to reach the given total momentum $Q$ (or the specified
quantum number $n$). 

It is worth noting that, for the case of the $k-\Lambda$ string solution
(i.e., the bound-state), we need to solve Eq. (\ref{eq:BetheAnsatzRootEquation})
for the complex-valued quasi-momenta $k_{1}=q+i\xi$ and $k_{2}=q-i\xi$,
with $\xi>0$. In addition, we need to specify $N-1$ integers $\{s_{j}\}$
($j=3,\cdots,N+1$) for the real-valued quasi-momenta $\{k_{j}\}$.
After some simple algebra, we find that the complex-valued quasi-momenta
satisfy the coupled equations (for a given $\Lambda$),
\begin{eqnarray}
\sin\left(q\right)\cosh\left(\xi\right) & = & \Lambda+u\frac{\sin\left(qL\right)}{\cosh\left(\xi L\right)-\cos\left(qL\right)}\equiv\tilde{\Lambda},\\
\cos\left(q\right)\sinh\left(\xi\right) & = & -u\frac{\sinh\left(\xi L\right)}{\cosh\left(\xi L\right)-\cos\left(qL\right)}\equiv-\tilde{u}.
\end{eqnarray}
For sufficiently large lattice size $L\gg1$, we may approximate $\tilde{\Lambda}\simeq\Lambda$
and $\tilde{u}\simeq u$ and then solve the coupled equations for
$q$ and $\xi$. This approximation gives a leading-order solution
for the values of $q$ and $\xi$. Numerically, we then improve $\tilde{\Lambda}$
and $\tilde{\xi}$, and \emph{iteratively} solve the coupled equations
until the convergence is reached. In the limit $\Lambda\rightarrow\pm\infty$,
we find that $q\rightarrow\pm\pi/2$ if $u<0$ or $q\rightarrow\mp\pi/2$
if $u>0$. 

\subsubsection{Attractive on-site interaction $U<0$}

It is straightforward to see that, for the solution with all real
quasi-momenta $\{k_{j}\}$,
\begin{eqnarray}
Q\left(\Lambda\rightarrow-\infty\right) & = & \frac{2\pi}{L}\sum_{j=1}^{N+1}s_{j},\\
Q\left(\Lambda\rightarrow+\infty\right) & = & \frac{2\pi}{L}\left[\sum_{j=1}^{N+1}s_{j}-\left(N+1\right)\right].
\end{eqnarray}
Therefore, for a given total momentum $Q=n(2\pi/L)$, where $-L/2<n\leq L/2$,
by tuning $\Lambda$ at most we can only find one value of $\Lambda$
that makes $Q=n(2\pi/L)$. For the $k-\Lambda$ string solution, we
find that,
\begin{eqnarray}
Q\left(\Lambda\rightarrow-\infty\right) & = & \frac{2\pi}{L}\left[-\frac{L}{2}+\sum_{j=3}^{N+1}s_{j}\right],\\
Q\left(\Lambda\rightarrow+\infty\right) & = & \frac{2\pi}{L}\left[+\frac{L}{2}+\sum_{j=3}^{N+1}s_{j}-\left(N-1\right)\right].
\end{eqnarray}
It is easy to see that, similarly at most we can only find one value
of $\Lambda$ that makes $Q=n(2\pi/L)$.

\subsubsection{Repulsive on-site interaction $U>0$}

For a repulsive on-site interaction, the situation is similar, although
the limiting values of the total momentum are reversed. However, one
needs to take care of the $k-\Lambda$ string solution. In this case,
by changing $\Lambda$ the value of $q$ has a jump at $\Lambda=0$,
if we restrict $q\in(-\pi,\pi]$. This unwanted jump can be readily
removed, if we add $2\pi$ to $q$ when $\Lambda<0$ or $q<0$.

\subsection{The set of quantum numbers $\{s_{j}\}$}

For large $L$ and $N$, the number of the regular Bethe wavefunctions
is exponentially large. We need to pick up the Bethe wavefunctions
with significant residue. A useful strategy is to classify the set
of integers $\{s_{j}\}$, according to the number of pseudo particle-hole
excitations, created by modifying a \emph{pseudo} Fermi sea configuration,
which always exists. 

\begin{figure}
\begin{centering}
\includegraphics[width=0.5\textwidth]{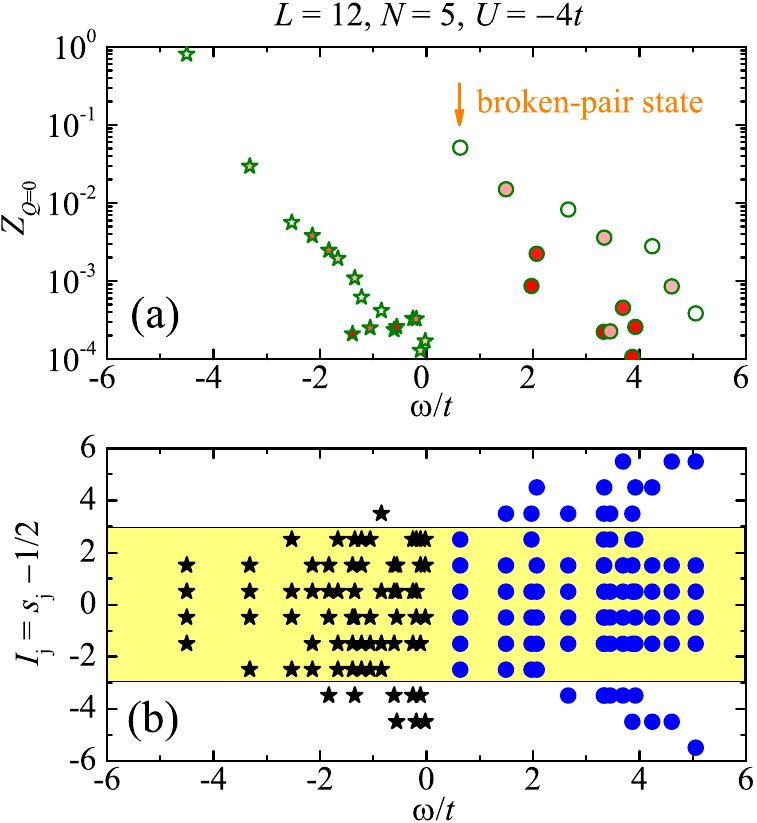}
\par\end{centering}
\caption{\label{fig2_states} Upper panel: Residues of different (i.e., $n$-th)
Bethe wavefunctions at zero total momentum ($Q=0$) as a function
of the energy $\omega=E_{N+1}^{(n)}\left(\{k_{j}\},\Lambda\right)-E_{\textrm{FS},N}$
at the quarter filling $\nu=(N+1)/L=0.5$ and at the attraction $U=-4t$.
We choose a small lattice size, i.e., $L=12$ and $N=5$, so the different
many-body states can be well separated, for a better illustration.
The stars and circles correspond to the $k-\Lambda$ solutions and
the all real-$k$ solutions, respectively. The color of the symbols
shows the value of the quasi-momentum $\Lambda$, which increases
as the color changes from blue to red. At zero momentum, all the states
are doubly degenerate, so we only include the states with $\Lambda\protect\geq0$.
The arrow emphasizes the broken-pair state, discovered by McGuire
with the Gaudin-Yang model \citep{McGuire1966}. Lower panel: Configurations
of the quantum numbers $\{I_{j}\}$ for different Bethe wavefunctions.
The yellow area highlights the pseudo Fermi sea of the broken-pair
state.}
\end{figure}

This pseudo Fermi sea configuration is given by the set $\{s_{j}\}=[-(N+1)/2+1,\cdots,-1,0,1,\cdots,(N+1)/2]$,
which characterize the ground state for repulsive on-site interactions
and a broken-pair state for attractive on-site interactions (see,
i.e., Fig. \ref{fig2_states}(b)). We refer to an integer $s_{H}$
in this pseudo Fermi sea set as a hole, and an integer $s_{P}$ out
of this set as a particle. More explicitly, we have $s_{H}\in[-(N+1)/2+1,\cdots,-1,0,1,\cdots,(N+1)/2]$
and $s_{P}\in[-L/2+1,\cdots,-(N+1)/2]\cup[(N+1)/2+1,\cdots,L/2]$.
In general, for the set $\{s_{j}\}$ named as $n$-particle-hole excitations,
we pick up $n$ particle values from $\{s_{P}\}$ and fill them into
(i.e., replace) the $n$ hole values of $\{s_{H}\}$. This idea applies
to the $k-\Lambda$ string solution as well. A slight modification
is that we need to discard two particle values in the $\{s_{j}\}$
set, to accommodate the pair of complex-valued quasi-momenta $k_{1}$
and $k_{2}=k_{1}^{*}$.

In our numerical calculations, we have tried to include the Bethe
wavefunctions with up to three pseudo particle-hole excitations, where
the sum-rule $\varrho_{s}=1$ is well satisfied. To have a large lattice
size and particle number, however, we often truncate to the level
of two pseudo particle-hole excitations. For example, at $L=100$
and $N=49$ and at the total momentum $Q=\pi$, we find that the numbers
of regular Bethe wavefunctions with one and two pseudo particle-hole
excitations are 2161 and 1182934, respectively. The inclusion of these
over-million quantum states leads to $\varrho_{s}>0.996$ at $U=4t$,
i.e., the sum-rule is satisfied within $0.4\%$ error.

\section{Results and discussions}

In the previous short Letter \citep{Hu2025}, we discussed the spectral
properties of the impurity particle at repulsive on-site interactions.
In this work, we would like to focus on attractive on-site interactions.
Due to the particle-hole transformation as discussed in Sec. IIC,
which transforms the interaction strength from $+U$ to $-U$, these
two situations have no difference in principle. However, the consideration
of an attractive on-site interaction seems to be favorable to understand
the polaron physics in one dimension. For example, it is well-known
that with attractive interactions, the impurity naturally has two
branches in its spectrum: a low-lying attractive Fermi polaron (or
Fermi singularity) and a high-lying excited repulsive Fermi polaron
(or Fermi singularity) \citep{Massignan2014,Wang2023AB,Knap2012,Wang2022PRL,Wang2022PRA}.
Therefore, it would be interesting to see how the two branches emerge
from the exact Bethe ansatz solution.

\begin{figure*}
\begin{centering}
\includegraphics[width=0.5\textwidth]{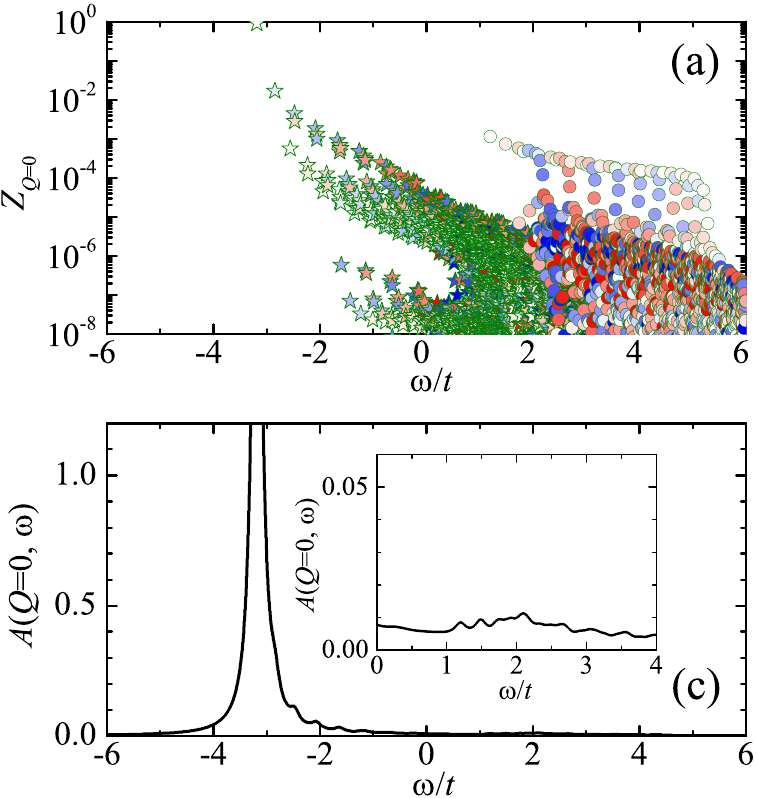}\includegraphics[width=0.5\textwidth]{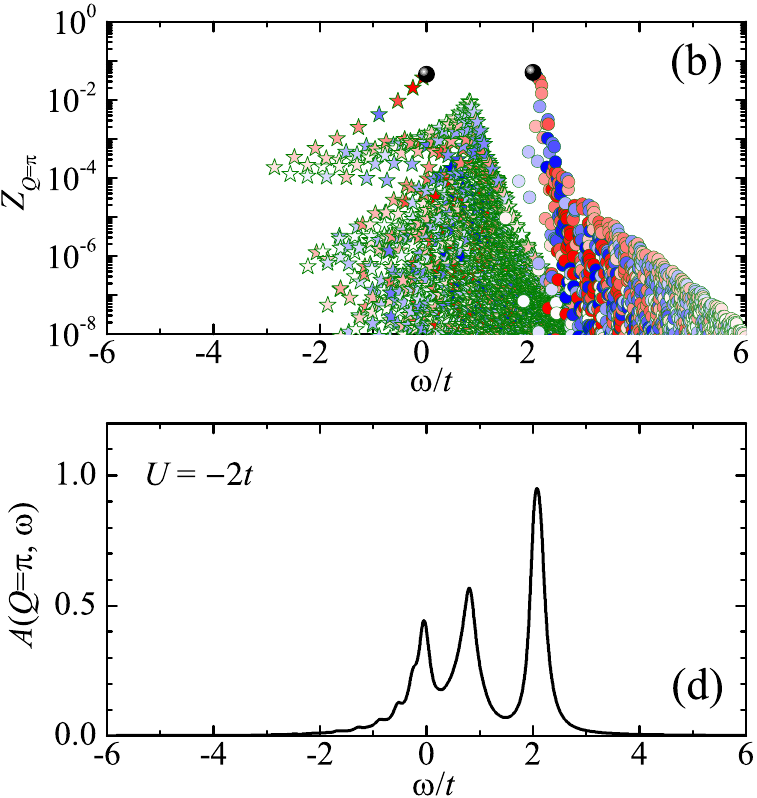}
\par\end{centering}
\caption{\label{fig3_uu05m} Residues (a and b) and the impurity spectral functions
(c and d, in units of $t^{-1}$) as a function of the energy $\omega=E_{N+1}^{(n)}\left(\{k_{j}\},\Lambda\right)-E_{\textrm{FS},N}$
at the quarter filling $\nu=(N+1)/L=0.5$ and at the attraction $U=-2t$.
In the left and right columns, we show the results at zero momentum
$Q=0$ and at the momentum $Q=\pi$, respectively. Here, we choose
$L=40$ and $N=19$. As in Fig. \ref{fig2_states}, in the upper panels,
the residues of the real-$k$ solutions and the $k-\Lambda$ solutions
are shown by circles and stars, respectively, with the color of symbols
indicating the value of the quasi-momentum $\Lambda$. The two black
dots in (b) show the residues of the spin-flip state (i.e., the right
dot) and of the $\eta$-pairing state (the left dot). The inset in
(c) highlights the repulsive branch of the spectral function starting
at $\omega\sim t$, contributed by all the real-$k$ solutions.}
\end{figure*}

Without loss of generality, we will primarily take a lattice size
$L=40$ and a filling factor $\nu=(N+1)/L=0.5$ (i.e., $N=19$), with
the inclusion of two pseudo particle-hole excitations. The sum rule
is satisfied at the level $\varrho_{s}>0.998$, so we believe that
the numerical convergence in choosing many-body states is well achieved.
The consideration of a relative small number of spin-up fermions $N$
follows the realistic experimental realization with ultracold atoms.
We note that, strictly speaking, in the thermodynamic limit the impurity
in one dimension might not behave like a polaron quasiparticle and
the impurity spectral function would be dominated by Fermi singularities
\citep{Castella1993,Hu2025}. However, experimentally, the characteristic
of Fermi singularities may hardly be identified, due to the small
number of particles in the Fermi bath.

\subsection{The configuration of Bethe wavefunctions}

Let us first discuss the configuration of different regular Bethe
wavefunctions, by considering a small lattice size $L=12$ and a small
particle number $N=5$ at an interaction strength $U=-4t$. Under
these parameters, the Bethe states are well separated in energy. In
Fig. \ref{fig2_states}, we present the set of quantum numbers $\{I_{j}=s_{j}-1/2\}$
of different Bethe states at zero total momentum (b) and the corresponding
form factors (a).

\begin{figure*}
\begin{centering}
\includegraphics[width=0.5\textwidth]{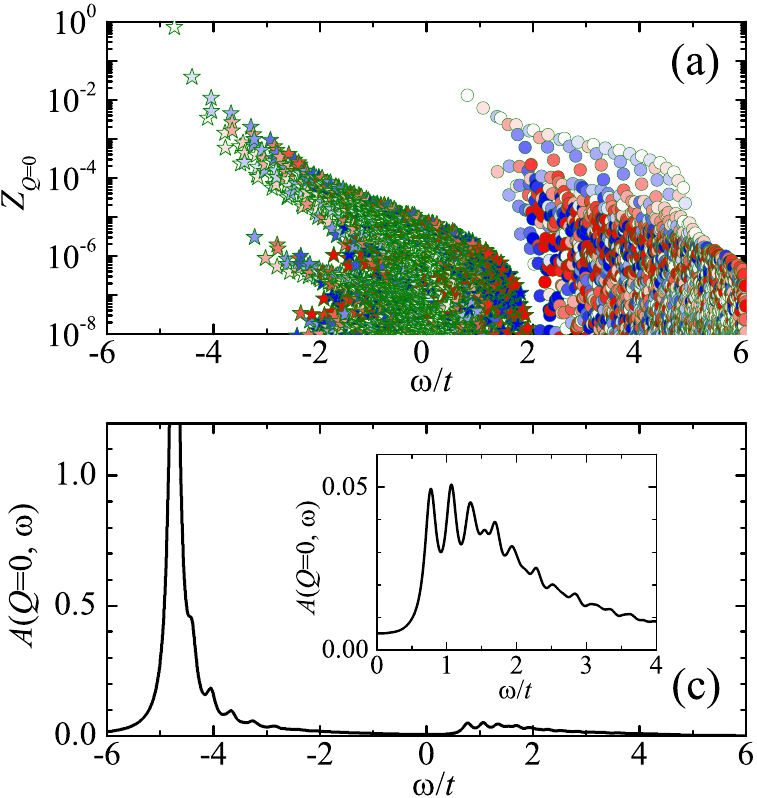}\includegraphics[width=0.5\textwidth]{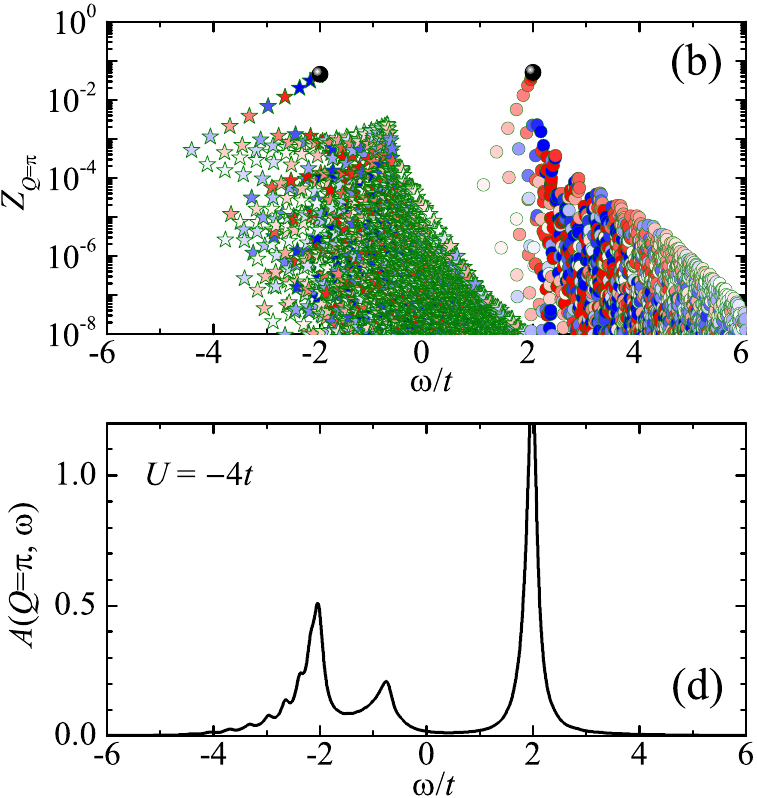}
\par\end{centering}
\caption{\label{fig4_uu10m} The same plots as in Fig. \ref{fig3_uu05m}, except
that the attractive interaction strength becomes $U=-4t$. Due to
the enhanced attraction, the repulsive branch of Fermi polarons at
zero momentum $Q=0$ becomes more significant, as highlighted in the
inset of Fig. \ref{fig4_uu10m}(c).}
\end{figure*}

At zero momentum $Q=0$, all the Bethe states are \emph{doubly} degenerate,
except some states with $\Lambda=0$ including the absolute ground
state. This degeneracy can be easily understood from the Bethe ansatz
equation Eq. (\ref{eq:BetheAnsatzRootEquation}), i.e., the equation
holds when we change the sign of $\{k_{j}\}$ and $\Lambda$: $k_{j}\rightarrow-k_{j}$
and $\Lambda\rightarrow-\Lambda$. As the constraint $Q=0=\sum_{j}k_{j}$
is not affected by this reverse operation, we would find two degenerate
states, with quasi-momentum $\Lambda\neq0$ and $-\Lambda$. In general,
the double degeneracy is lost once the total momentum $Q$ becomes
nonzero. An exception is the case with the total momentum $Q=\pi$.
As the momentum $\pi$ is related to the momentum $-\pi$ by the reciprocal
lattice momentum $2\pi$, the constraint on the total momentum $Q=\pi=\sum_{j}k_{j}$
is again not affected by the reverse operation $k_{j}\rightarrow-k_{j}$.
Therefore, the double degeneracy is recovered at the total momentum
$Q=\pi$. As a result of the double degeneracy at $Q=0$, for clarity
in Fig. \ref{fig2_states}(b) we show only the configuration $\{I_{j}\}$
of the states with the quasi-momentum $\Lambda\geq0$. The configuration
of the degenerate state is simply given by the set $\{\tilde{I}_{j}=-I_{j}\}$.

From their configurations as a function of energy, we find that the
all real-$k$ states and the $k-\Lambda$ string states are roughly
separated by the on-site attraction $U=-4t$. For the attractive on-site
interaction, the all real-$k$ state with the lowest energy, known
as the broken-pair state \citep{McGuire1966}, has the configuration
of a pseudo Fermi sea. Typically, the states with a large energy has
large particle numbers $\left|s_{j,P}\right|$ or $\left|I_{j,P}\right|$
out of the pseudo Fermi sea.

The separation of the two kinds (or branches) of states can also been
clearly seen from the form factor shown in Fig. \ref{fig2_states}(a).
In each branch, the form factor generally decreases with increasing
energy. In the lower branch of the $k-\Lambda$ string states, the
absolute ground state has the largest residue, which is almost unity.
In the upper branch of the all real-$k$ states, the broken-pair has
the largest residue. However, the magnitude of its residue is significantly
smaller.

\subsection{The form factor and spectral function}

In Fig. \ref{fig3_uu05m}, we report the form factors (upper panels)
and spectral functions (lower panels) of a realistic situation with
$L=40$ and $N=19$ at the on-site attraction $U=-2t$. At zero total
momentum $Q=0$, as given in Fig. \ref{fig3_uu05m}(a), the form factor
shows a similar characteristic as in Fig. \ref{fig2_states}(a). The
large residue of the absolute ground state leads to a sharp peak at
the ground state energy $\omega=E_{N+1}^{(0)}\left(\{k_{j}\},\Lambda\right)-E_{\textrm{FS},N}\sim-3t$
(see, i.e., Fig. \ref{fig3_uu05m}(c)), where the superscript ``0''
indicates the ground state. At the small number of particles $N=19$,
we may treat this peak as a polaron quasiparticle, as the contribution
from the nearby discrete excited states is negligible. For example,
the first excited state has a residue at about $0.01$, so the sharp
peak is essentially contributed by the ground state only. By increasing
$L$ and $N$, the situation will be different, since the nearby excited
states will become densely distributed and their residues will eventually
render the sharp peak into a Fermi singularity \citep{Castella1993,Hu2025}.
We note that, at the weak attraction $U=-2t$, we can hardly identify
the existence of the high-energy branch contributed from the all real-$k$
states. At the energy of the broken-pair state, i.e., $\omega\sim t$,
we can only see a small, very broad background distribution, as highlighted
in the inset of Fig. \ref{fig3_uu05m}(c), whose magnitude is consistent
with the residue of the broken-pair state (i.e., $Z\sim0.001$).

\begin{figure*}
\begin{centering}
\includegraphics[width=0.5\textwidth]{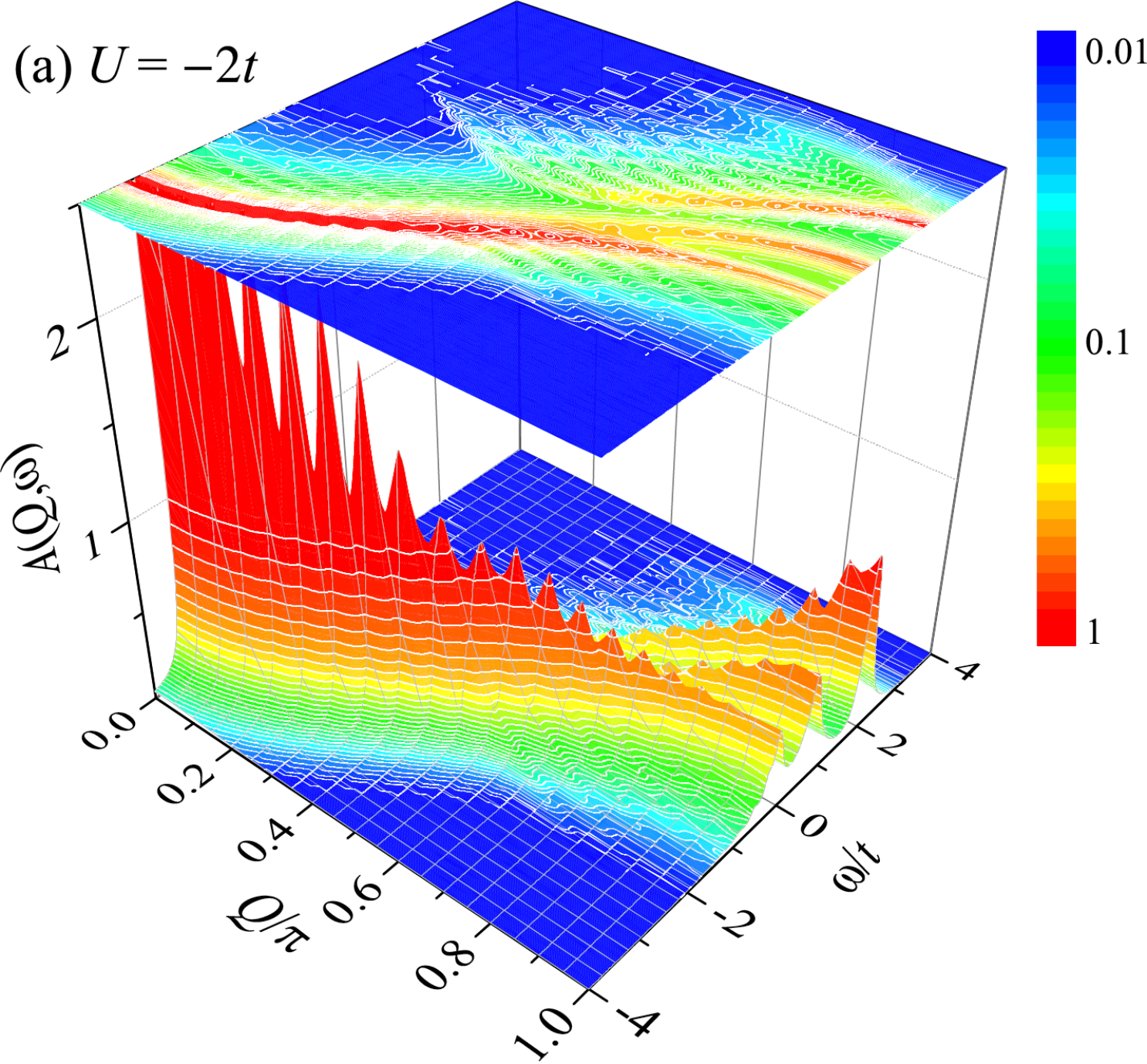}\includegraphics[width=0.5\textwidth]{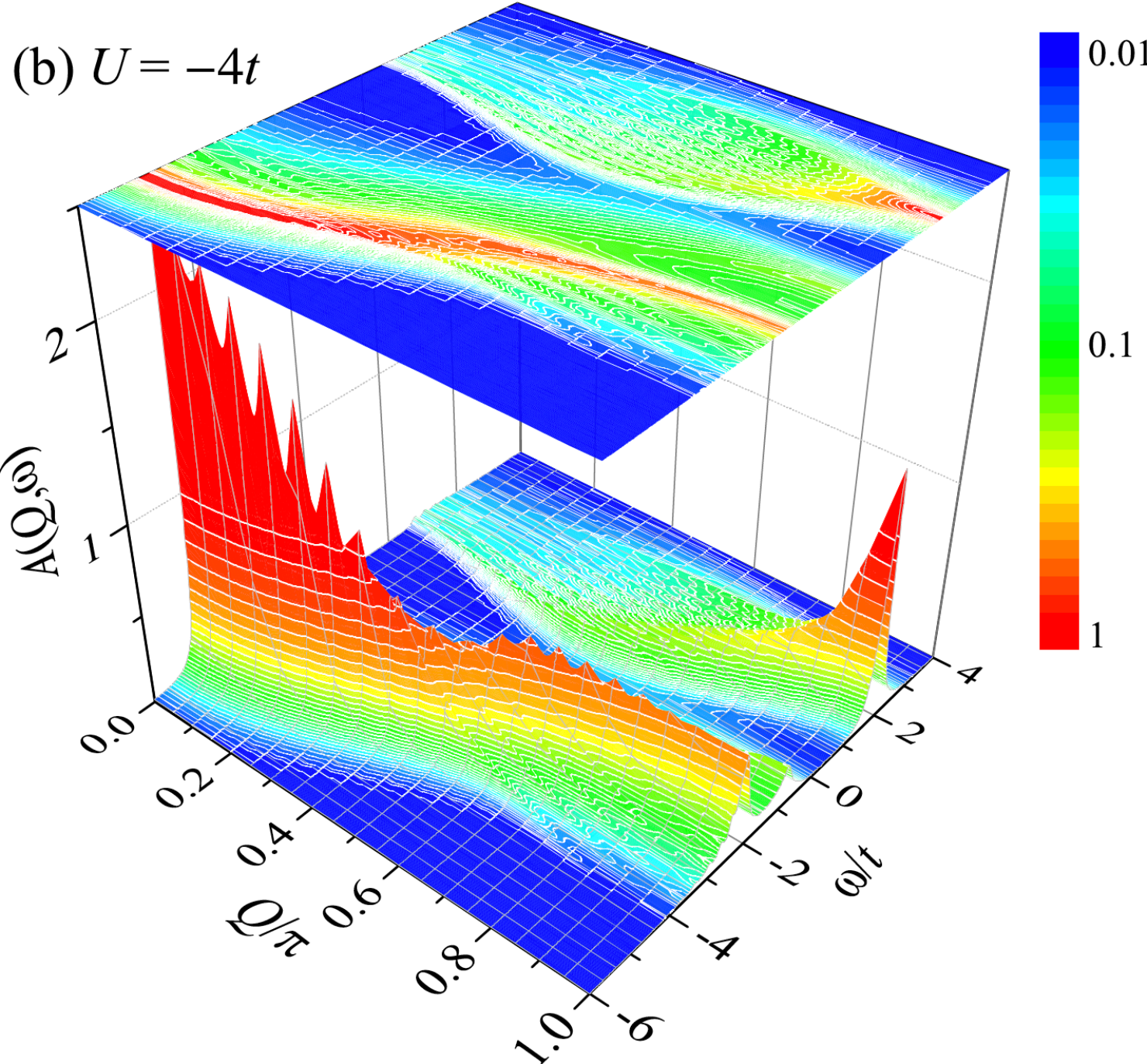}
\par\end{centering}
\caption{\label{fig5_akw3d} Three-dimensional plots of the spectral function
$A(Q,\omega)$, in units of $t^{-1}$ as indicated by the color map,
at the interaction strengths $U=-2t$ (a) and $U=-4t$ (b). The corresponding
two-dimensional contour plots are also shown at the top of the figure.
Here, we take the lattice size $L=40$ and the number of spin-up fermions
$N=19$.}
\end{figure*}

At the total momentum $Q=\pi$, as shown in Fig. \ref{fig3_uu05m}(b),
we find the behavior of the form factors as a function of the energy
becomes very different. At this sector, compared with the $Q=0$ case,
we find three distinct features. First, the lowest energy state still
locates at a similar energy $\omega\sim-3t$. However, its residue
$Z\sim10^{-4}$ becomes significantly smaller. By increasing energy,
we find a series of the $k-\Lambda$ string states whose residue rapidly
increases. This series stops at the $\eta$-pairing state $\left|\eta\right\rangle $
(see the left black dots in Fig. \ref{fig3_uu05m}(b)), which precisely
locates at the energy $\omega_{\eta}=-2t\cos Q+U=0$. Second, at a
bit higher energy $\omega\sim t$, we find a cluster of the $k-\Lambda$
string states, whose residues are notable (i.e., $Z\sim0.01$). The
number of states in the cluster is not small. Finally, we can also
see clearly a series of the all real-$k$ states, connecting smoothly
to the spin-flip state $\left|\xi\right\rangle $ at the energy $\omega_{\xi}=-2t\cos Q=2t$,
which is indicated by the right black dot in Fig. \ref{fig3_uu05m}(b).
Remarkably, the three features in the form factors yield three peaks
in the spectral function, as illustrated in Fig. \ref{fig3_uu05m}(d).
It is worth emphasizing that the middle peak II, contributed by the
cluster of the $k-\Lambda$ string states, would be a \emph{true}
polaron quasiparticle, which will survive in the thermodynamic limit
\citep{Hu2025}. In contrast, the first peak (located near $\omega_{\eta}=0$)
and the third peak (located near $\omega_{\xi}=2t$) will eventually
turn into Fermi singularities with increasing $L$ and $N$ \citep{Hu2025}.
This coexistence of the Fermi singularities and the polaron quasiparticle
in the thermodynamic limit could be a unique feature of Fermi polarons
in one dimension.

In Fig. \ref{fig4_uu10m}, we present the results for the form factors
and spectral functions at a larger attractive interaction, $U=-4t$.
By increasing attraction, at the total momentum $Q=0$, we find that
the high-lying repulsive branch in the spectral function become more
significant (see the inset of Fig. \ref{fig4_uu10m}(c)). At the total
momentum $Q=\pi$, the three features in the form factors become more
pronounced and well separated. Moreover, in the spectral function
the amplitude of the middle polaron peak II becomes smaller. We have
tried other values of on-site attractive interaction strengths and
find that the polaron peak II will completely disappear at sufficiently
large attractions.

In closing this subsection, let us discuss the total momentum $Q$-dependence
of the spectral function. As reported in Fig. \ref{fig5_akw3d}, we
show the three-dimensional plots of the spectral function at the two
on-site attractions, $U=-2t$ (a) and $U=-4t$ (b). The projected
two-dimensional contour plots are given at the top of the figures.
By increasing the total momentum $Q$, the low-lying branch contributed
from the $k-\Lambda$ string states and the high-lying branch from
the all real-$k$ states give rise to the well-known lower and upper
Hubbard bands, respectively. These two bands are clearly visible at
the attraction $U=-4t$. At a smaller attraction $U=-2t$, we also
see an additional middle band split from the lower Hubbard band, responsible
for the middle polaron quasiparticle peak that we emphasized in Fig.
\ref{fig3_uu05m}(d). This middle band also exists at $U=-4t$ at
the total momentum $Q\sim\pi$. However, it becomes very shallow and
eventually disappear at large attractions.

\begin{figure*}
\begin{centering}
\includegraphics[width=0.33\textwidth]{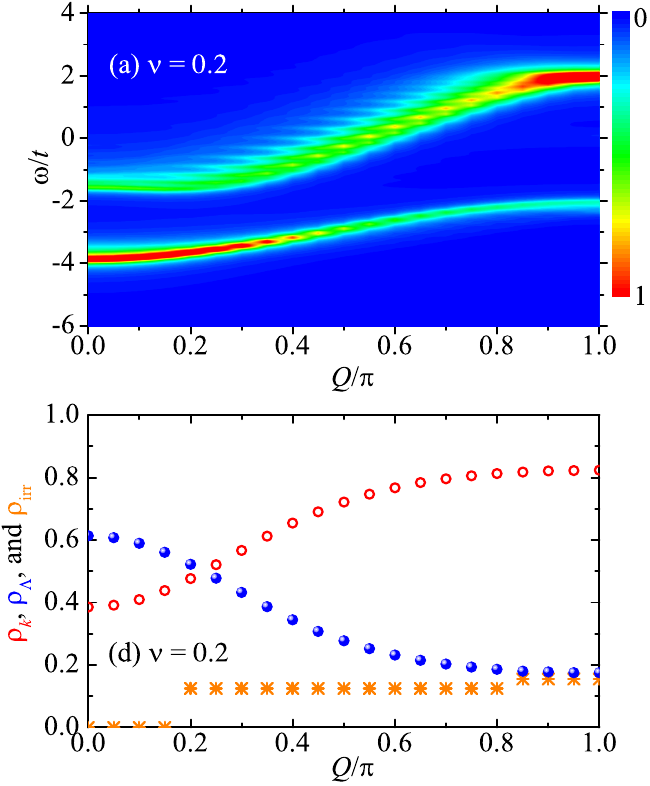}\includegraphics[width=0.33\textwidth]{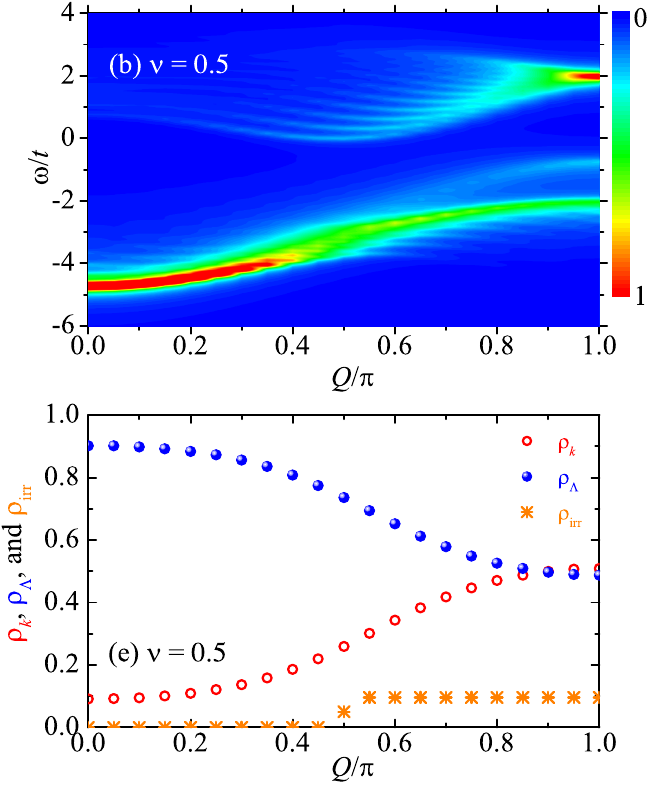}\includegraphics[width=0.33\textwidth]{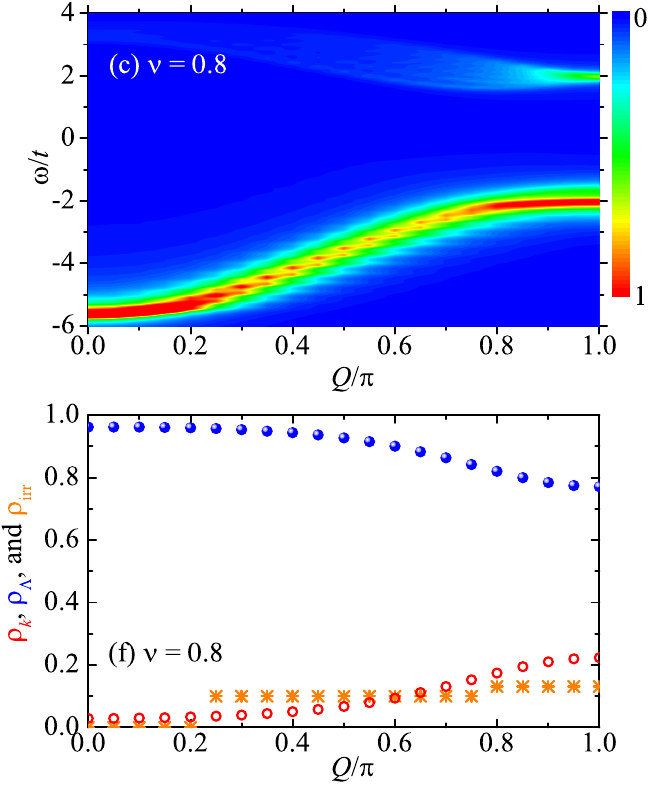}
\par\end{centering}
\caption{\label{fig6_akw2d} Upper panels: two-dimensional contour plots of
the spectral function $A(Q,\omega)$, in units of $t^{-1}$ as indicated
by the color map, at various filling factors $\nu=0.2$ (a), $\nu=0.5$
(b) and $\nu=0.8$ (c). Lower panels (d, e, f): the corresponding
sum of residues, contributed by the all real-$k$ solutions and the
spin-flip state (red open circles, $\varrho_{k}$), by the $k-\Lambda$
string solutions and the $\eta$-pairing state (blue solid circles,
$\varrho_{\Lambda}$). Note that, $\varrho_{k}+\varrho_{\Lambda}=1$
following the sum rule. We also show the sum of the residues of the
irregular spin-flip and $\eta$-pairing states as $\varrho_{\textrm{irr}}$
(orange stars). Here, we consider the interaction strength $U=-4t$,
the lattice size $L=40$, and the number of spin-up fermions $N=\nu L-1$.}
\end{figure*}

\subsection{The dependence on the filling factor}

Let us finally consider the filling factor $\nu$-dependence of the
impurity spectral function. In Fig. \ref{fig6_akw2d}, we report the
two-dimensional contour plots of the spectral function at three filling
factors $\nu=0.2$ (a), $\nu=0.5$ (b) and $\nu=0.8$ (c), and at
the on-site attraction $U=-4t$. The corresponding sums of the residues
$\varrho_{\Lambda}$, $\varrho_{k}$ and $\varrho_{\textrm{irr}}$,
contributed from different quantum states are provided in the lower
panels of the figure. 

We find that $\varrho_{\Lambda}$ and $\varrho_{k}$ provide a useful
measure of the relative strength of the low-lying and high-lying Hubbard
bands, as a function of the total momentum $Q$. By increasing $Q$,
the spectral weight of the low-lying Hubbard band (i.e., $\varrho_{\Lambda}$)
decreases and the spectral weight of the high-lying Hubbard band (i.e.,
$\varrho_{k}$) increases. At the zero momentum ($Q=0$), the low-lying
Hubbard band typically has more spectral weight than the high-lying
band. At the total momentum $Q=\pi$, we find that the values of $\varrho_{\Lambda}$
and $\varrho_{k}$ can be well approximated by $\nu$ and $1-\nu$,
respectively.

\section{Conclusions and outlooks}

In conclusions, we have derived an analytic expression of the form
factor for an impurity moving in a non-interacting Fermi bath, based
on the exactly solvable one-dimensional Hubbard model. This analytic
expression allows us to exactly calculate the spectral function of
Fermi polarons in one-dimensional optical lattices with lattice sizes
$L\sim100$ and to address the finite-size effect in the spectral
properties by varying the lattice size. The same expression of the
form factor could also be used to investigate the non-trivial quantum
dynamics of the impurity, such as quantum flutter in optical lattices
\citep{Mathy2012,Knap2014,Gamayun2018}.

The current derivation of the form factor strongly relies on the fact
that the many-body Bethe wavefunctions can be rewritten into a Slater
determinant. A similar technique might be used to solve the problem
of an impurity moving in an interacting Bose gas in one dimension,
which is described by the bosonic Gaudin-Yang model with extreme spin-population
imbalance \citep{Li2003}. This would provide us some exact results
to better understand the spectral properties of one-dimensional Bose
polarons.

In future studies, it would also be interesting to consider multiple
fermionic impurities interacting with the one-dimensional Fermi bath.
The finite number or finite density of impurities will introduce a
pseudo Fermi sea of spinons for the spin degree of freedom \citep{Essler2005}.
As a result, we should be able to see spinon excitations in the spectral
properties and the appearance of the intriguing spin-charge separation
in the single-particle spectral function \citep{Kohno2010}.

\section{Statements and Declarations}

\subsubsection{Ethics approval and consent to participate }

Not Applicable.

\subsubsection{Consent for publication }

Not Applicable.

\subsubsection{Availability of data and materials }

The data generated during the current study are available from the
contributing author upon reasonable request. 

\subsubsection{Competing interests}

The authors have no competing interests to declare that are relevant
to the content of this article. 

\subsubsection{Funding}

This research was supported by the Australian Research Council's (ARC)
Discovery Program, Grants Nos. DP240100248 (X.-J.L.) and DP240101590
(H.H.). 

\subsubsection{Authors' contributions }

All the authors equally contributed to all aspects of the manuscript.
All the authors read and approved the final manuscript. 

\subsubsection{Acknowledgements }

See funding support.

\subsubsection{Authors' information}

Xia-Ji Liu, Centre for Quantum Technology Theory, Swinburne University
of Technology, Melbourne 3122, Australia, Email: xiajiliu@swin.edu.au

Hui Hu, Centre for Quantum Technology Theory, Swinburne University
of Technology, Melbourne 3122, Australia, Email: hhu@swin.edu.au

\end{document}